\documentclass[twocolumn]{aastex631}
\usepackage{amsmath}
\usepackage{multirow}

\newcommand{\blue}[1]{\textcolor{blue}{#1}}
\newcommand{\black}[1]{\textcolor{black}{#1}}

\begin{document}

\title{Revisiting radio variability of the blazar 3C 454.3}

\author[0000-0002-3960-5870]{Ashutosh Tripathi}
\affiliation{George P.\ and Cynthia Woods Mitchell Institute for Fundamental Physics and Astronomy,\\ Texas A\&M University, College Station, TX 77843-4242, USA}
\affiliation{Xinjiang Astronomical Observatory, CAS, 150 Science-1 Street, Urumqi 830011, People's Republic of China}
\affiliation{Department of Physics, Southern Methodist University, 3215 Daniel Avenue, 
Dallas, Texas 75205, USA}
\correspondingauthor{Ashutosh Tripathi}
\email{ashutosh31tripathi@yahoo.com}

\author[0000-0002-9331-4388]{Alok C.\ Gupta}
\affiliation{Aryabhatta Research Institute of Observational Sciences (ARIES), Manora Peak, Nainital 263001, India}

\author[0000-0001-5785-7038]{Krista Lynne Smith}
\affiliation{George P.\ and Cynthia Woods Mitchell Institute for Fundamental Physics and Astronomy,\\ Texas A\&M University, College Station, TX 77843-4242, USA}
\affiliation{Department of Physics, Southern Methodist University, 3215 Daniel Avenue, 
Dallas, Texas 75205, USA}

\author[0000-0002-1029-3746]{Paul J.\ Wiita}
\affiliation{Department of Physics, The College of New Jersey, 2000 Pennington Rd., Ewing, New Jersey 08628-0718, USA}

\author[0000-0003-2483-2103]{Margo F.\ Aller}
\affiliation{Department of Astronomy, University of Michigan, 1085 S. University Avenue, Ann Arbor, Michigan 48109, USA}

\author[0000-0002-3839-3466]{Alexandr E.\ Volvach}
\affiliation{Radio Astronomy Laboratory, Crimean Astrophysical Observatory, Katsiveli, Crimea}

\author[0000-0002-0393-0647]{Anne L\"ahteenm\"aki} 
\affiliation{Aalto University Mets\"ahovi Radio Observatory, Mets\"ahovintie 114, 02540 Kylm\"al\"a, Finland}
\affiliation{Aalto University Department of Electronics and Nanoengineering, P.O. BOX 15500, FI-00076 Aalto, Finland}

\author[0000-0003-1945-1840]{Hugh D.\ Aller}
\affiliation{Department of Astronomy, University of Michigan, 1085 S. University Avenue, Ann Arbor, Michigan 48109, USA}

\author[0000-0003-1249-6026]{Merja Tornikoski} 
\affiliation{Aalto University Mets\"ahovi Radio Observatory, Mets\"ahovintie 114, 02540 Kylm\"al\"a, Finland}

\author[0000-0001-6157-003X]{Larisa N.\ Volvach}
\affiliation{Radio Astronomy Laboratory, Crimean Astrophysical Observatory, Katsiveli, Crimea}


\begin{abstract}
We examine lengthy radio light curves of the flat spectrum radio galaxy 3C 454.3 for possible quasi-periodic oscillations (QPOs). The data used in this work were collected at five radio frequencies, 4.8, 8.0, 14.5, 22.0, and 37.0 GHz between 1979--2013 as observed at the University of Michigan Radio Astronomical Observatory, Crimean Astrophysical Observatory, and Aalto University Mets{\"a}hovi Radio Observatory. We employ generalized Lomb-Scargle periodogram and weighted wavelet transform analyses \textcolor{black}{to search for periodicities in these light curves}. We confirm a QPO period of $\sim$ 2000 day to be at least 4$\sigma$ significant using both  methods at all five radio frequencies between 1979 and 2007, after which a strong flare changed the character of the light curve. We also find a $\sim$~600 day period which is at least 4$\sigma$ significant, but only in the 22.0 and 37.0 GHz light curves. We briefly discuss physical mechanisms capable of producing such variations. 
\end{abstract}


\section{Introduction} \label{sec:intro}
\noindent
Blazars are radio-loud active galactic nuclei (AGNs) possessing relativistic jets pointed almost toward the observer \citep{1995PASP..107..803U}. 
Due to this small inclination angle,  relativistic effects are important and result in substantially magnified observed emissions, such that the jet emission dominates the overall \black{observed fluxes from}  
blazars \citep{1995PASP..107..803U}. Blazars exhibit extraordinary flux, spectral and polarization variability throughout the electromagnetic (EM) spectrum \citep[e.g.][and references therein]{2010Natur.463..919A,2015ApJ...807...79H,2017MNRAS.472..788G,2017Natur.552..374R,2023MNRAS.526.4502R,2022Natur.609..265J,2022Natur.611..677L,2023ApJ...948L..25P,2023ApJ...953L..28M}.
BL Lacerate objects (BL Lacs) and flat spectrum radio quasars (FSRQs) are collectively called  blazars. In the composite optical/UV spectrum, BL Lacs show featureless or very weak emission lines (equivalent width EW $\leq$ 5\AA) \citep{1991ApJ...374...72S,1996MNRAS.281..425M} whereas FSRQs have prominent emission lines \citep{1978PhyS...17..265B,1997A&A...327...61G}.  
The spectral energy distributions (SEDs) of blazars 
show a double-humped composition. The lower energy hump peaks between infrared and X-ray bands and is a synchrotron emission that originates from relativistic electrons in the jet. The high energy hump peaks in $\gamma-$rays and are  \textcolor{black}{commonly} explained by inverse Compton (IC) radiation  \citep{2007Ap&SS.309...95B,2018A&A...616L...6G}. \\
\\
Periodic, or more properly, quasi-periodic, oscillations (QPOs), have been observed frequently in the light curves (LCs) of stellar-mass black hole (BH) and neutron star (NS) binaries \citep{2006ARA&A..44...49R}. But the LCs of AGNs across the entire EM spectrum  
are mostly non-periodic, with stochastic variations that can be attributed to instabilities in the accretion disks or jets \citep[see][and references therein]{2018A&A...616L...6G,2021MNRAS.501.5997T,2024MNRAS.527.9132T}. However, in the last 15 years or so occasional detections of QPOs in different EM bands with diverse periods have been reported in several blazars \citep[e.g.][and references therein]{2009ApJ...690..216G,2009A&A...506L..17L,2013MNRAS.436L.114K,2015ApJ...813L..41A,2016AJ....151...54S,2017A&A...600A.132S,2018NatCo...9.4599Z,2019MNRAS.487.3990B,2020A&A...642A.129S,2021MNRAS.501...50S,2021MNRAS.501.5997T,2022MNRAS.510.3641R,2022MNRAS.513.5238R,2022Natur.609..265J,2024MNRAS.527.9132T,2024MNRAS.528.6608T} and other classes of AGNs \citep[e.g.][and references therein]{2008Natur.455..369G,2014MNRAS.445L..16A,2015MNRAS.449..467A,2016ApJ...819L..19P,2018A&A...616L...6G}. \\ 
\\
The QPOs detected in stellar-mass black holes in X-ray binary systems and in Seyfert galaxies, likely due to disk phenomena, have oscillation frequencies that are inversely proportional to their mass \citep{2004ApJ...609L..63A, 2015ApJ...798L...5Z}. This captivating relation seems to be valid for both stellar mass BHs and supermassive black holes (SMBHs) in Seyfert galaxies and quasars whose emission is not jet-dominated. This suggests that the mechanism responsible for such periodicity / quasi-periodicity likely is similar in both types of central objects. Hence the detection of periodicity in an AGN LC could be used to determine the object's mass using this astounding relation. As these QPO signals almost certainly originate in the inner part of the accretion disk they could also be used to study the gravitational effects of the central object on its surroundings. Various  models have been proposed to explain this phenomenon \citep[e.g.][]{1999A&A...349.1003T, 2001ApJ...559L..25W, 2009MNRAS.397L.101I}, but the physical mechanism responsible for these features remains uncertain. \\  
\\
3C 454.3 is a \black{luminous} FSRQ at redshift $z =$ 0.859 \citep{1980ApJS...43...57H}, and \black{was} the  brightest blazar in 0.1 -- 10 GeV $\gamma-$rays \textcolor{black}{during the outburst in \black{2009--2011  \citep{2011ApJ...736L..38V}}}.  
Its SMBH mass has been estimated in the range of (0.5 -- 2.3) $\times$ 10$^{9} {\rm M}_{\sun}$ \citep[e.g.][and references therein]{2017MNRAS.472..788G,2019A&A...631A...4N}. Several dedicated simultaneous multi-wavelength (MW) observation campaigns have been made of this source to understand its incredible and peculiar variability across the whole EM spectrum \citep[e.g.][and references therein]{2006A&A...445L...1F, 2006A&A...449L..21P, 2006A&A...453..817V, 2007A&A...464L...5V, 2009A&A...504L...9V, 2006A&A...456..911G, 2007A&A...473..819R, 2008A&A...485L..17R, 2008A&A...491..755R, 2011A&A...534A..87R, 2008ApJ...676L..13V, 2009ApJ...690.1018V, 2010ApJ...712..405V, 2011ApJ...736L..38V, 2009ApJ...699..817A, 2009ApJ...707.1115D, 2010ApJ...715..362J, 2013ApJ...773..147J, 2010ApJ...716L.170P, 2010ApJ...721.1383A, 2012AJ....143...23G, 2012ApJ...758...72W, 2017MNRAS.472..788G, 2019ApJ...887..185S}. 
The first radio observation of this source was reported in \citep{1967Natur.213..977K}. After that, 3C 454.3 was extensively observed in radio wavebands \citep{1969ApJ...158..849M, 1971AJ.....76..537F, 1975ApJ...201..256S, 1980ApJ...236..714P, 1981AJ.....86..371P}. \textcolor{black}{\citet{2013MNRAS.436.3341Z} studied the radio emissions from the jet of this source \black{using very large baseline array (VLBI) observations (MOJAVE Collaboration)} and concluded that 3C 454.3 exhibits  \black{a} possible large-scale, ordered magnetic field component present hundreds of parsecs from its launching location.  \citet{2019ApJ...875...15W} found the size of the emission region to be of the order of $10^{15}$ cm using the optical data in the $R$ band \black{taken} by various ground-based telescopes.} \citet{2021A&A...648A..27V} analyzed the radio observations taken at Simeiz (RT-22) and claimed that this source could possibly be the most massive SMBH binary system.  \\
\\


\begin{figure*}[t]
\hspace{-2.5cm}\includegraphics[scale=0.55]{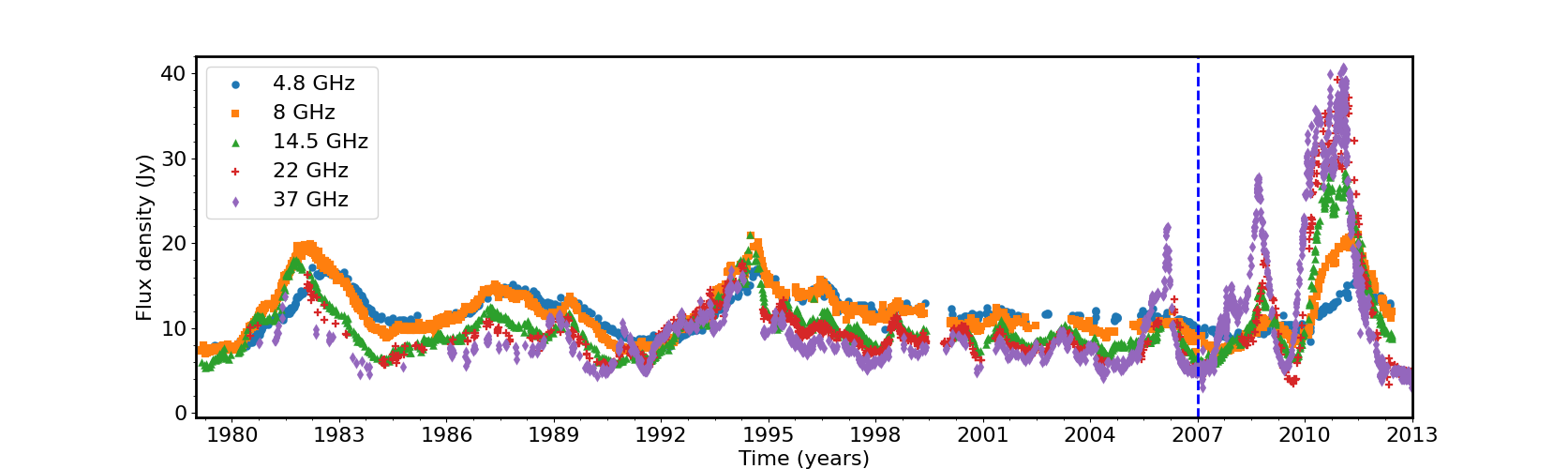}\\
\caption{Long-term light curves of 3C 454.3 at 4.8, 8.0, 14.5, 22, and 37  GHz during 1979--2013. The dashed blue line divides the light curves into Segments 1 and 2. }
\end{figure*}\label{fig:lc}

\begin{figure*}[t]

\hspace{-1.0cm}\includegraphics[scale=0.28]{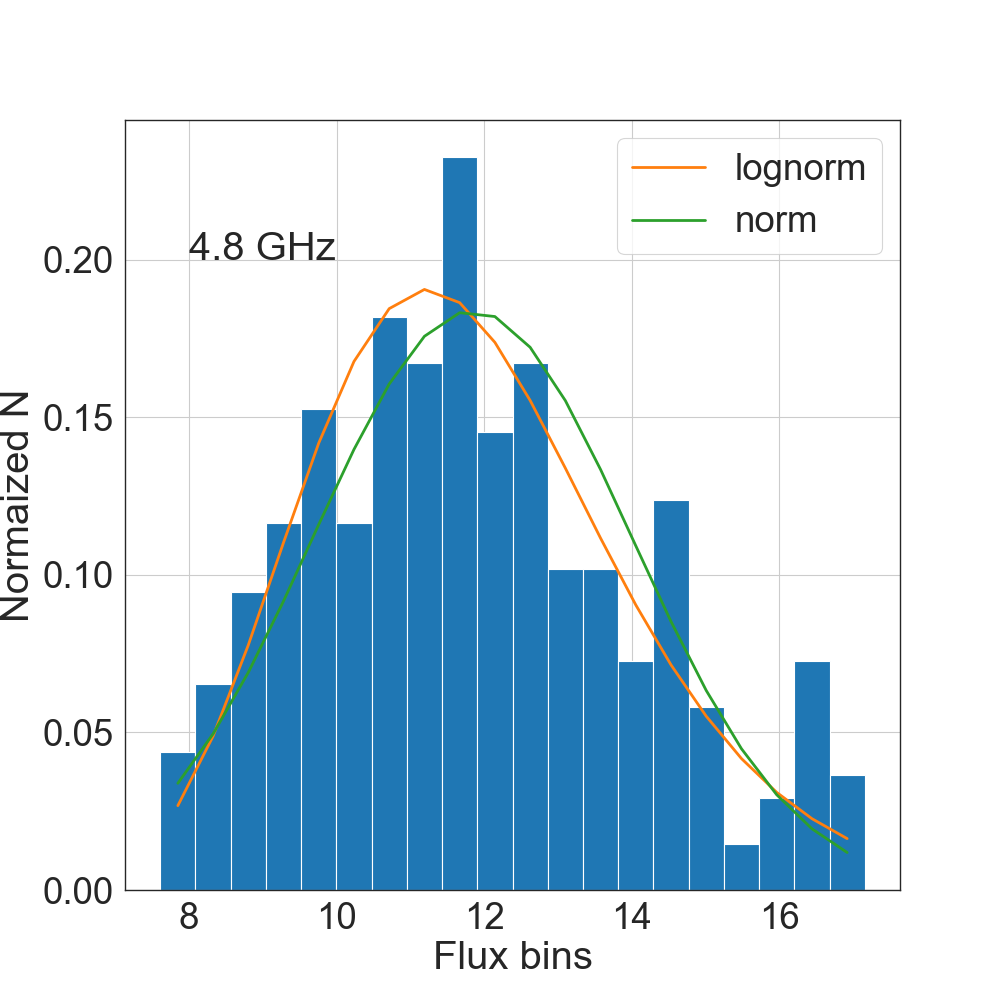}\hspace{-0.8cm} \includegraphics[scale=0.28]{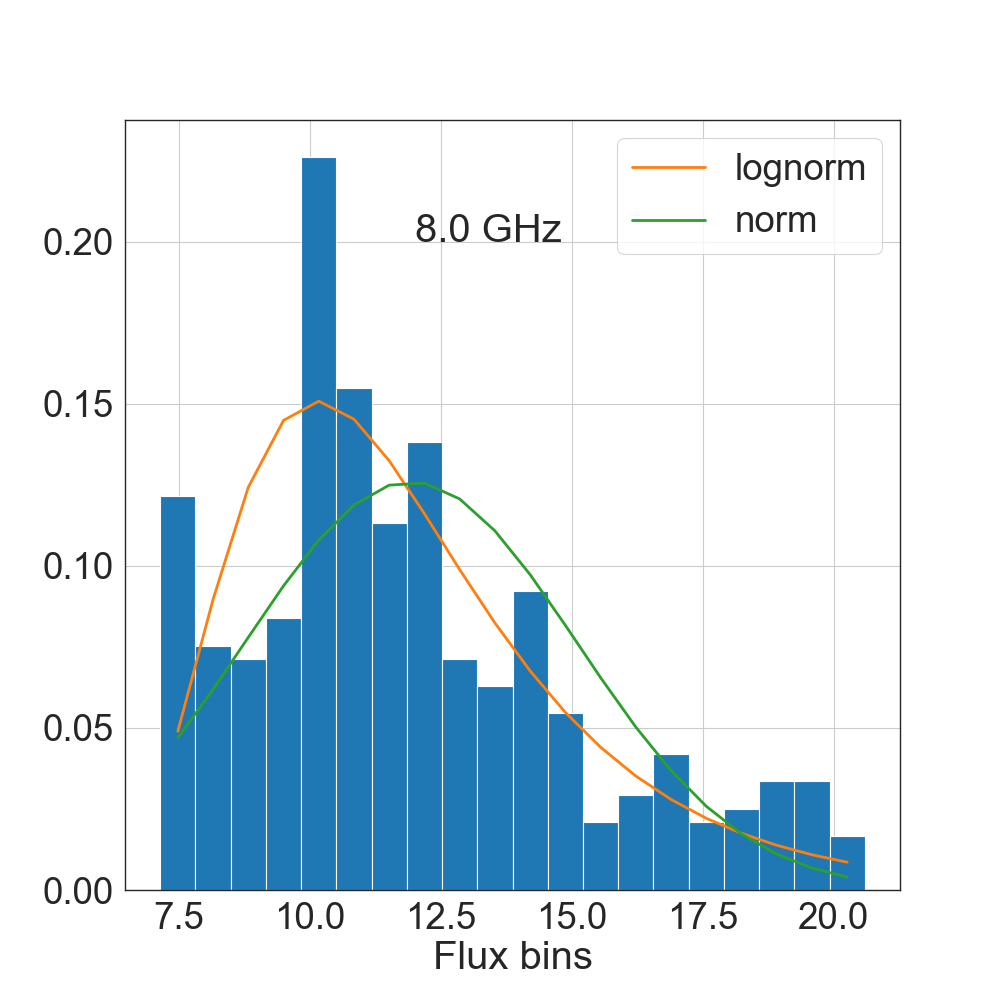}\includegraphics[scale=0.28]{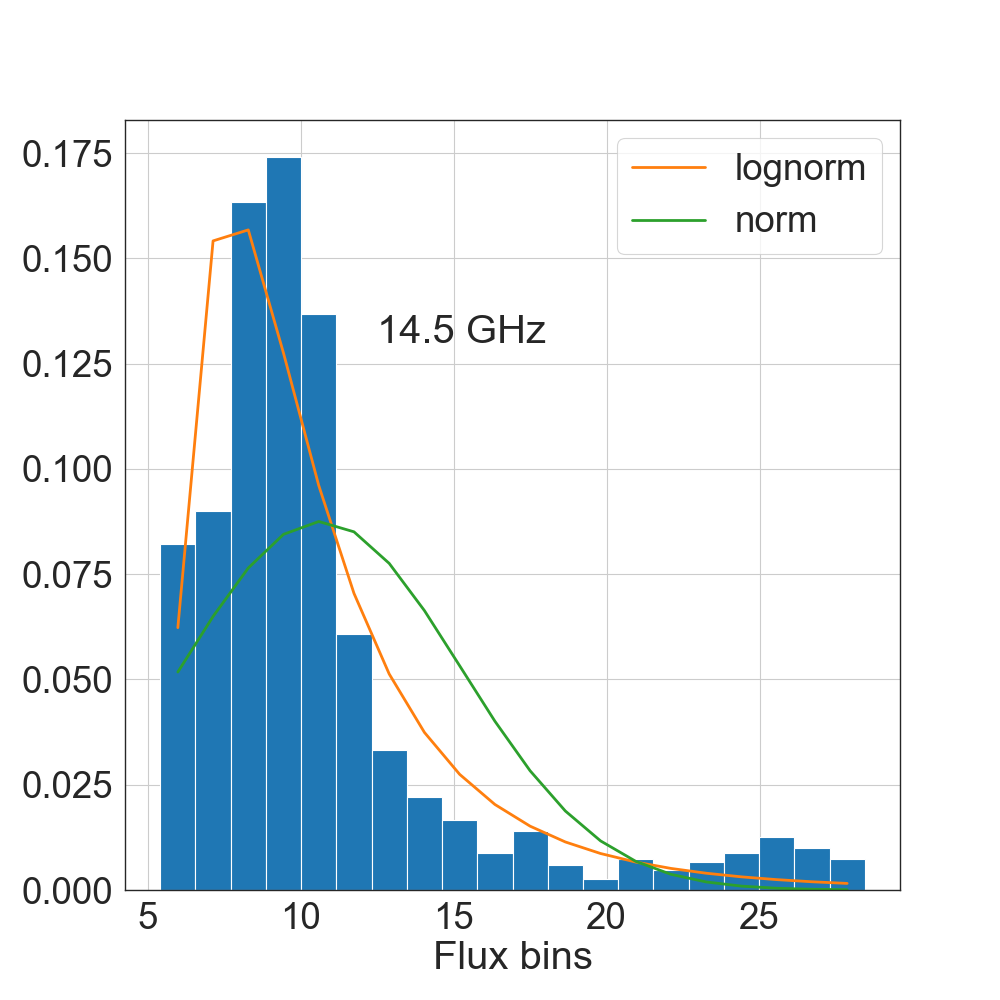}\\

\centering\includegraphics[scale=0.28]{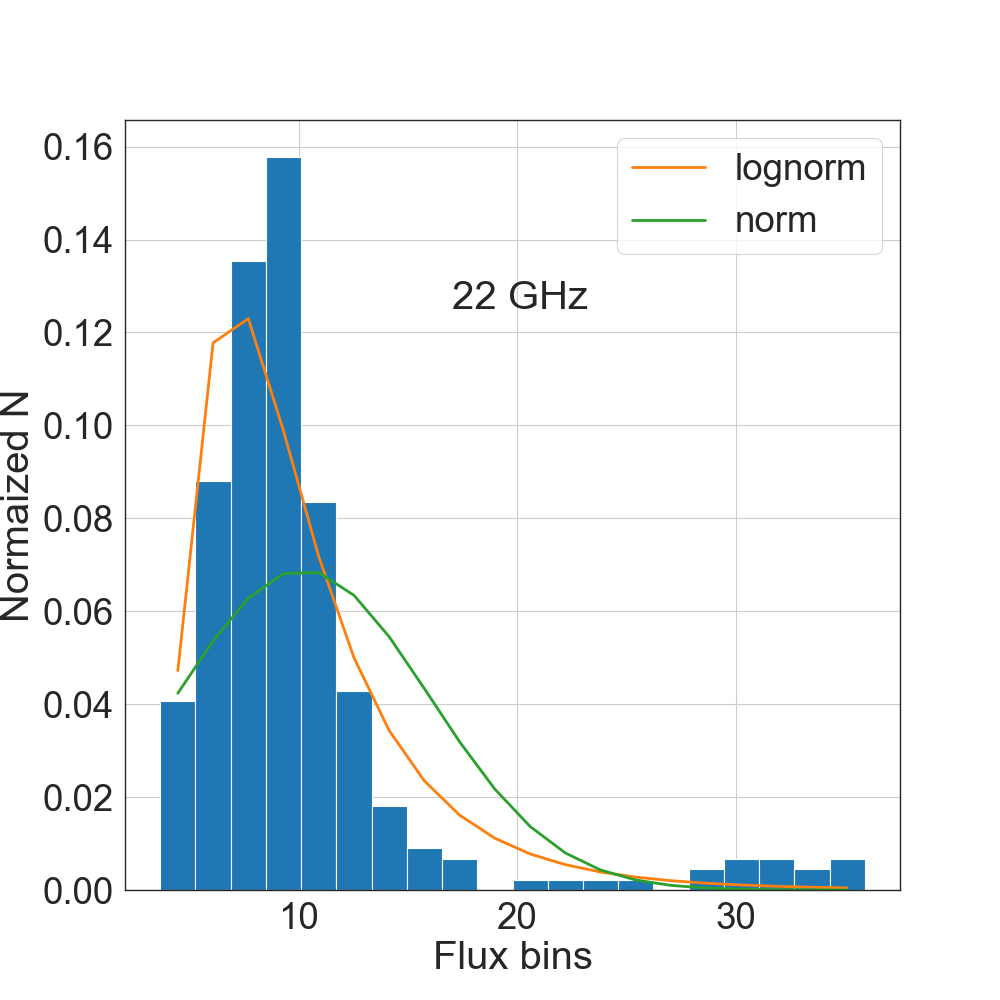}\hspace{-0.7cm} \includegraphics[scale=0.28]{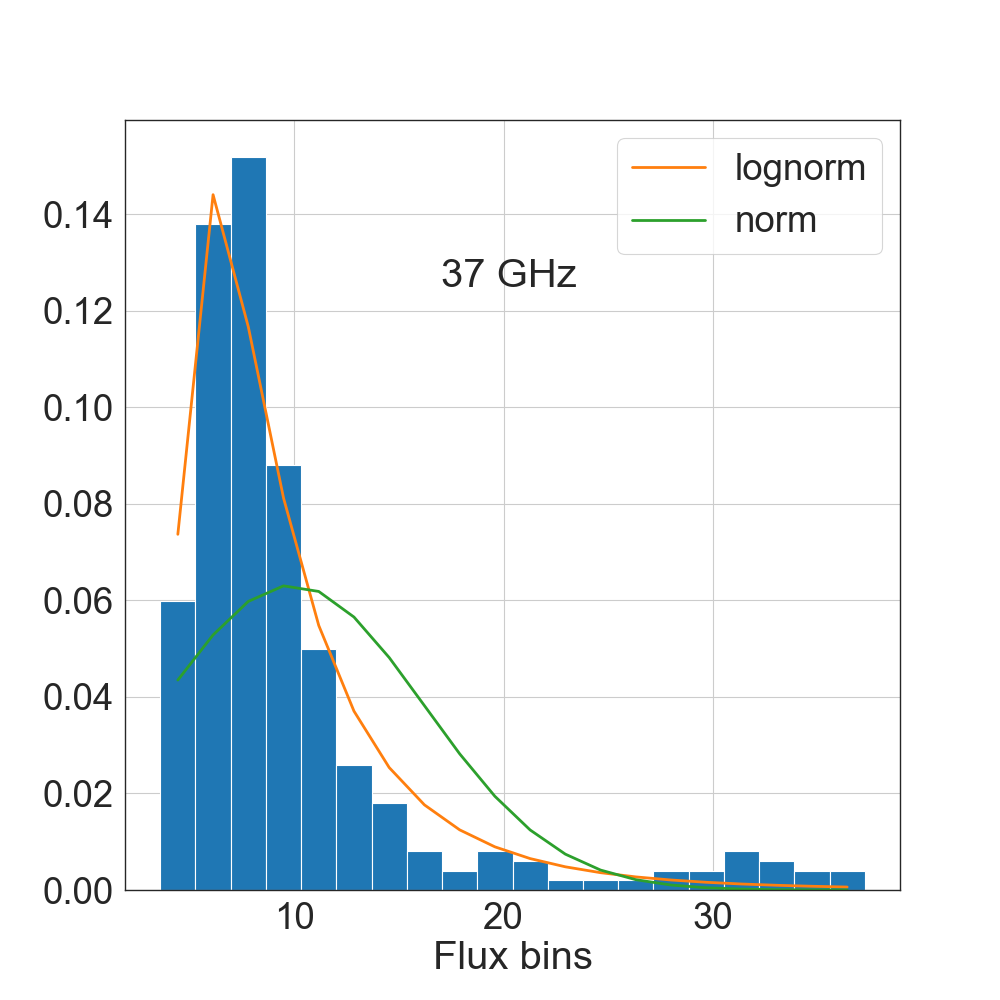}\\

\caption{Flux distribution histograms for the light at radio frequencies at 4.8, 8.0, 14.5, 22.0, and 37.0 GHz and their \black{best} fits with log-normal and normal distribution.}
\end{figure*}\label{fig:lc}

\noindent
Periodicities of the order of a few days to a few years have been claimed for this source in various wavebands. \citet{2004A&A...419..485C} reported a QPO in the range of 6.0 -- 6.5 yr in the radio data taken from 1970 to 1999 by UMRAO and Mets{\"a}hovi radio telescopes in the frequencies 4.8, 8, 14.5, 22 and 37 GHz. \citet{2007A&A...462..547F} also claimed periods in the range of 4.5 -- 13.6 years at the radio frequencies of 4.8, 8.0, and 14.8 GHz taken at UMRAO and Mets{\"a}hovi radio telescopes. In the optical waveband, QPOs in the range of 0.83 -- 12.1 years have been suggested \citep{1969Natur.221..755L, 1988AJ.....95..374W, 2000AcApS..20...11S, 2022ApJS..262...43Y}. \citet{2019RAA....19..142F} analyzed the optical g, r, and i band observations and claimed the presence of a $\sim$~100 min periodicity. 
\citet{2021MNRAS.501...50S} reported the quasi-periodicity of 47d in the $\gamma$-ray light curve of 3C 454.3, along with a hint of an optical QPO of the same \textcolor{black}{duration}.\\
\\
\\
In this paper, we have revisited the search for QPOs in the radio LCs of 3C 454.3 collected at the frequencies of 4.8, 8.0, 14.5, 22.0, and 37.0 GHz during 1979--2013. Aside from confirming the presence of a $\sim$~2000 day QPO we also note \black{that} a periodicity of $\sim$~600 days  appears to be present in higher frequency radio  observations. However, this \black{possible} QPO becomes less significant after 2007, \black{most likely} because of new strong flaring activity. In Sec.~\ref{sec:obs} we describe the radio observations used in this work and the data analysis methods used to analyze them. Sections ~\ref{sec:res} and 4 outlines the results obtained for these 5 radio LCs using those data analysis techniques. We  also describe the detection of a $\sim$~600 d signal in the light curves at 22 and 37 GHz and the effect of flares on periodogram calculations.  We  discuss some plausible physical models and give our conclusions in Sec.~\ref{sec:dis}.

\begin{figure*}[t]
\centering
\includegraphics[scale=0.35]{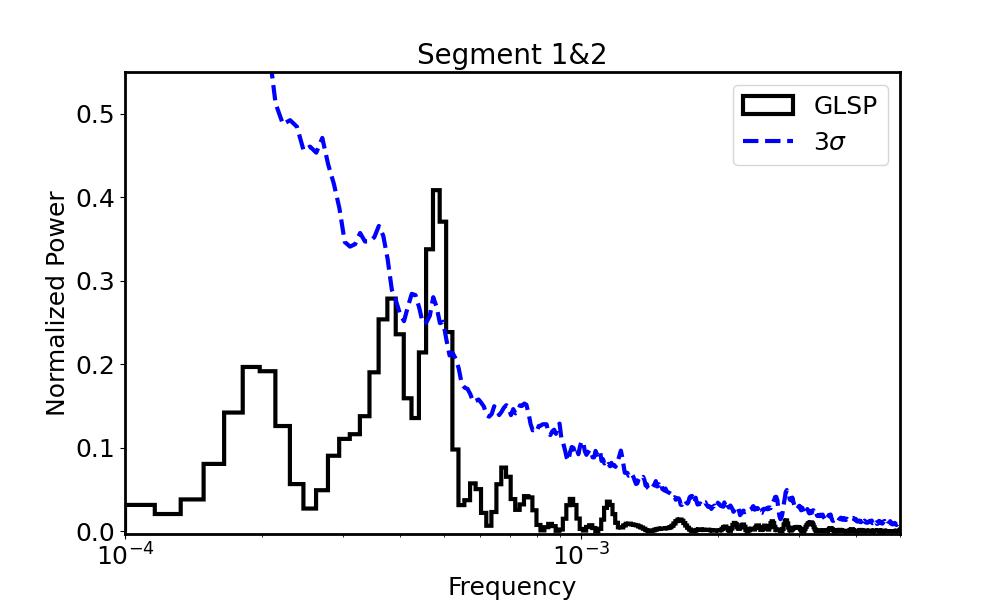}\includegraphics[scale=0.35]{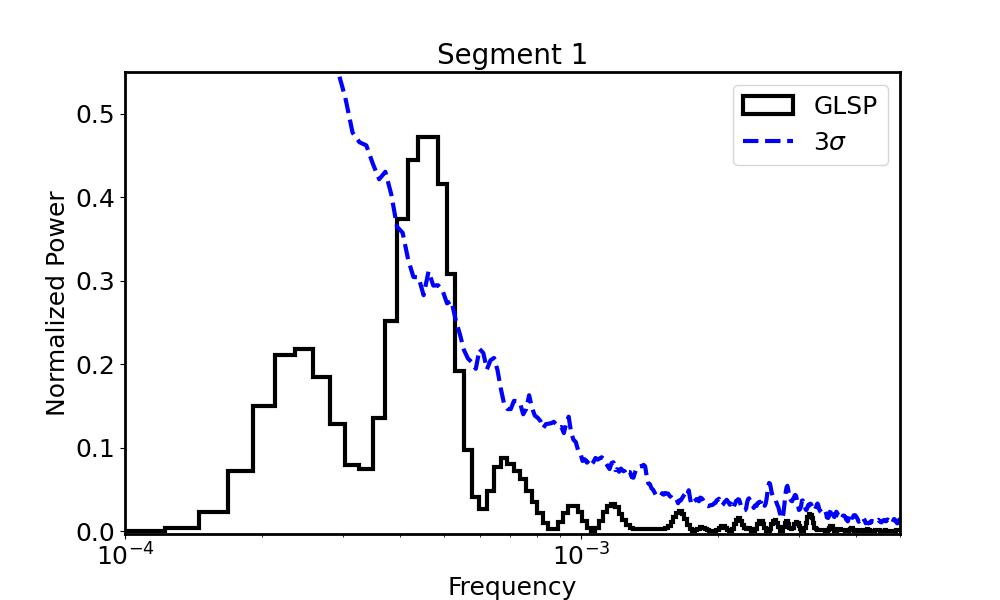}\\
\vspace{-1.0cm}
\includegraphics[angle=90, scale = 0.3]{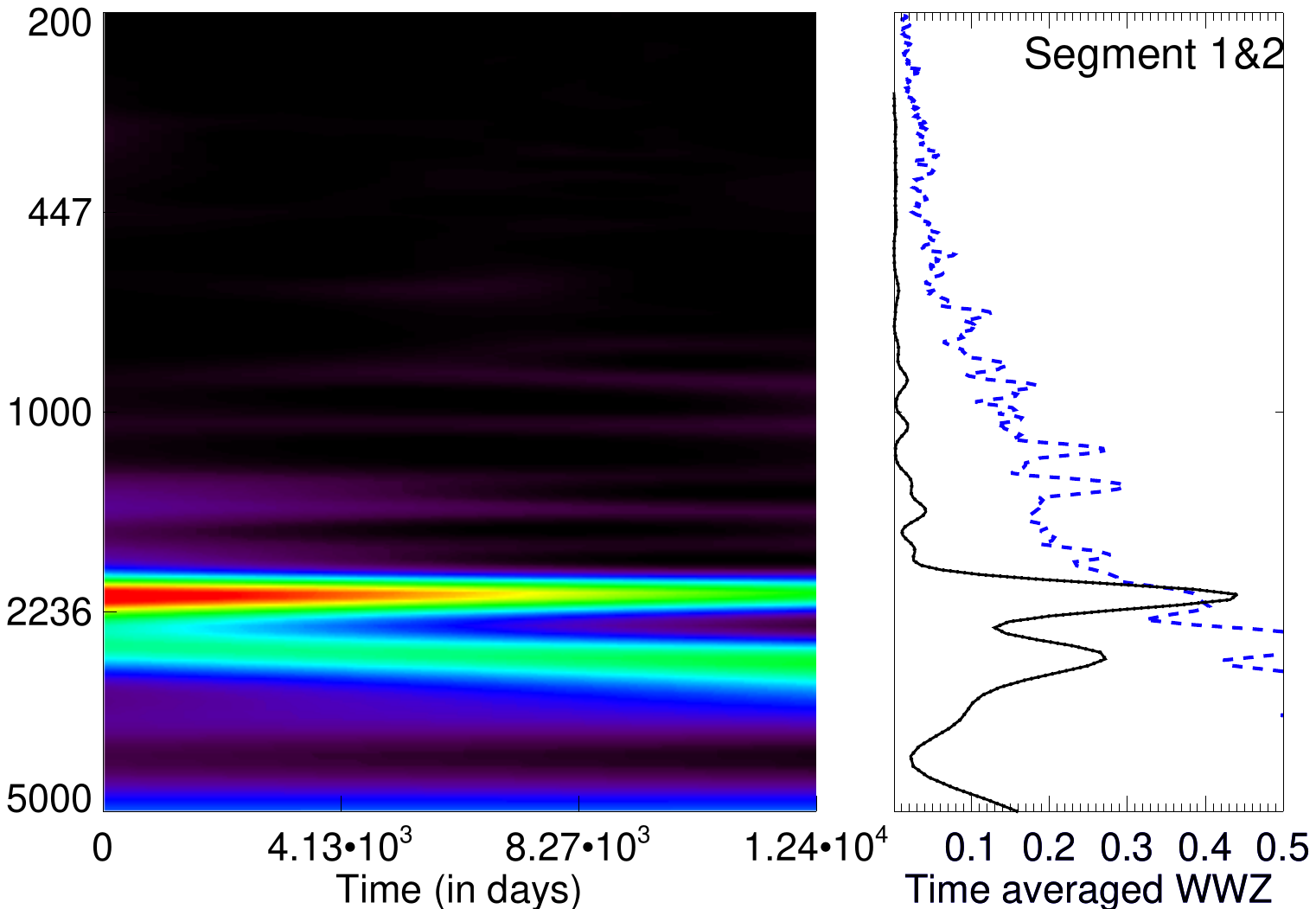}\includegraphics[angle=90, scale = 0.3]{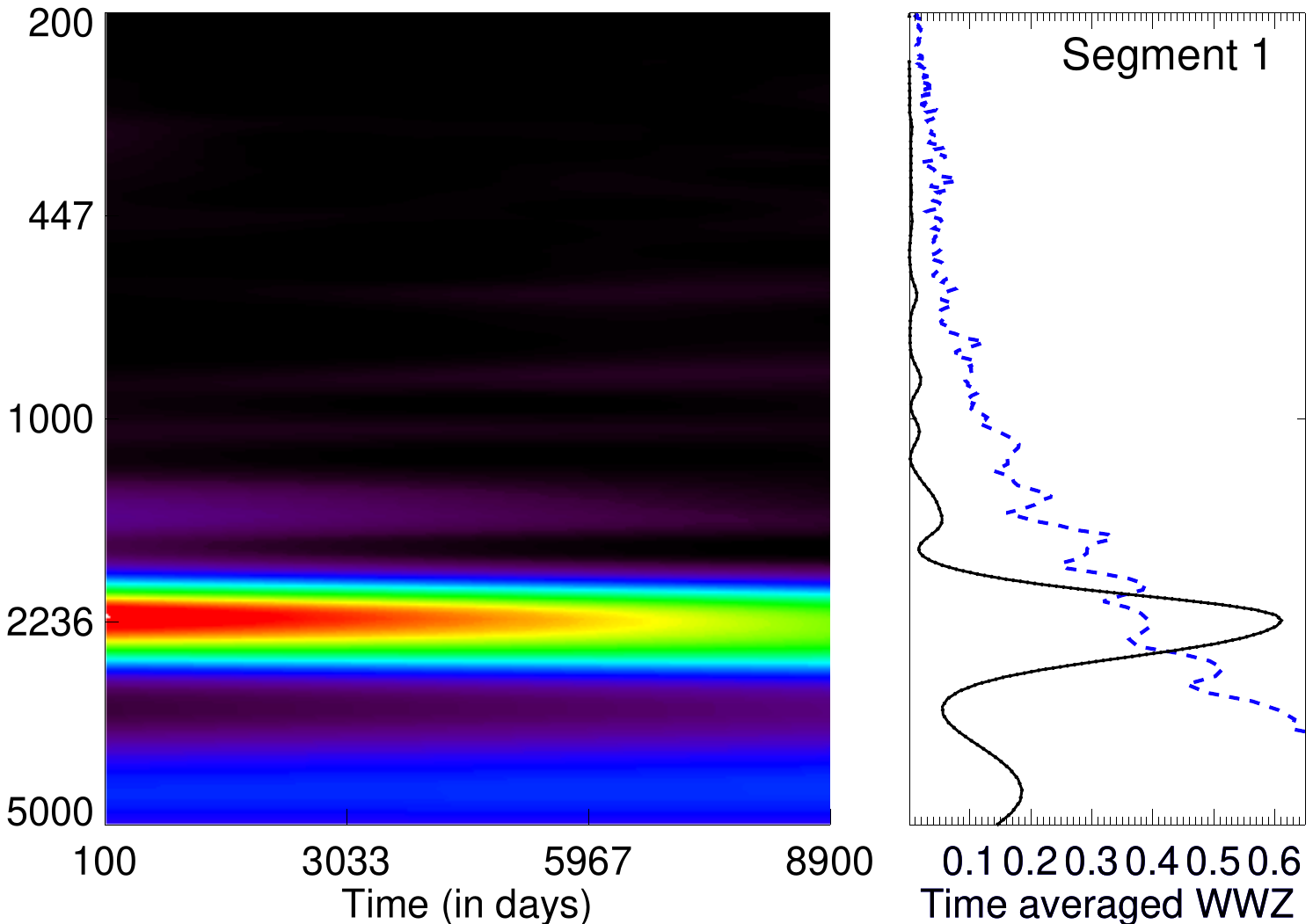}\\
\caption{ {\it Upper panel}: Generalized Lomb-Scargle periodogram for the whole light curve (left) and Segment 1 (right) at 4.8 GHz. \textcolor{black}{The blue dashed curves respectively denote 3$\sigma$ significance using a broken power law as the underlying red noise model}. {\it Lower panel}: Wavelet analysis for the entire light curve (left) and Segment 1 (right). In each panel, the left plot is the color-color diagram showing the WWZ power, where red denotes the most concentrated power and the power decreases toward violet and black. The right plot shows the time-averaged WWZ along with 3$\sigma$ significance given by the dashed blue curve. }
\end{figure*}\label{fig:5}

\begin{figure*}[t]
\centering
\includegraphics[scale=0.35]{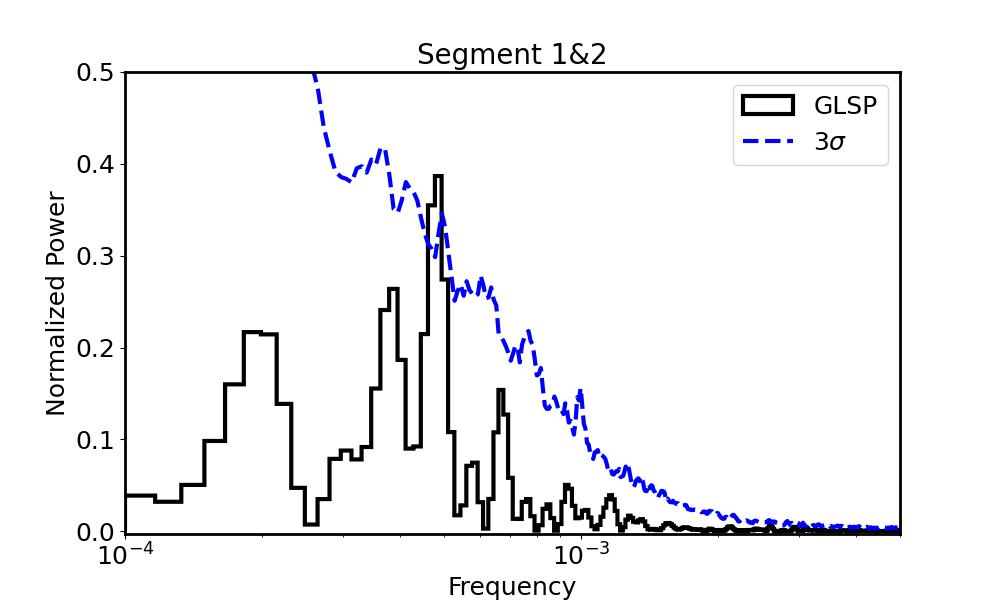}\includegraphics[scale=0.35]{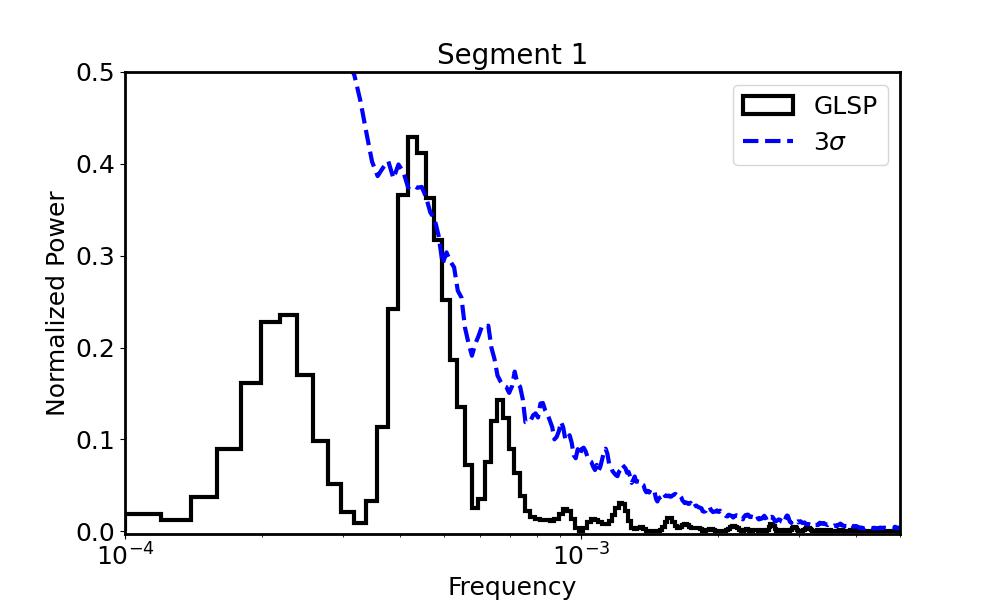}\\
\vspace{-1.0cm}
\includegraphics[angle=90, scale = 0.3]{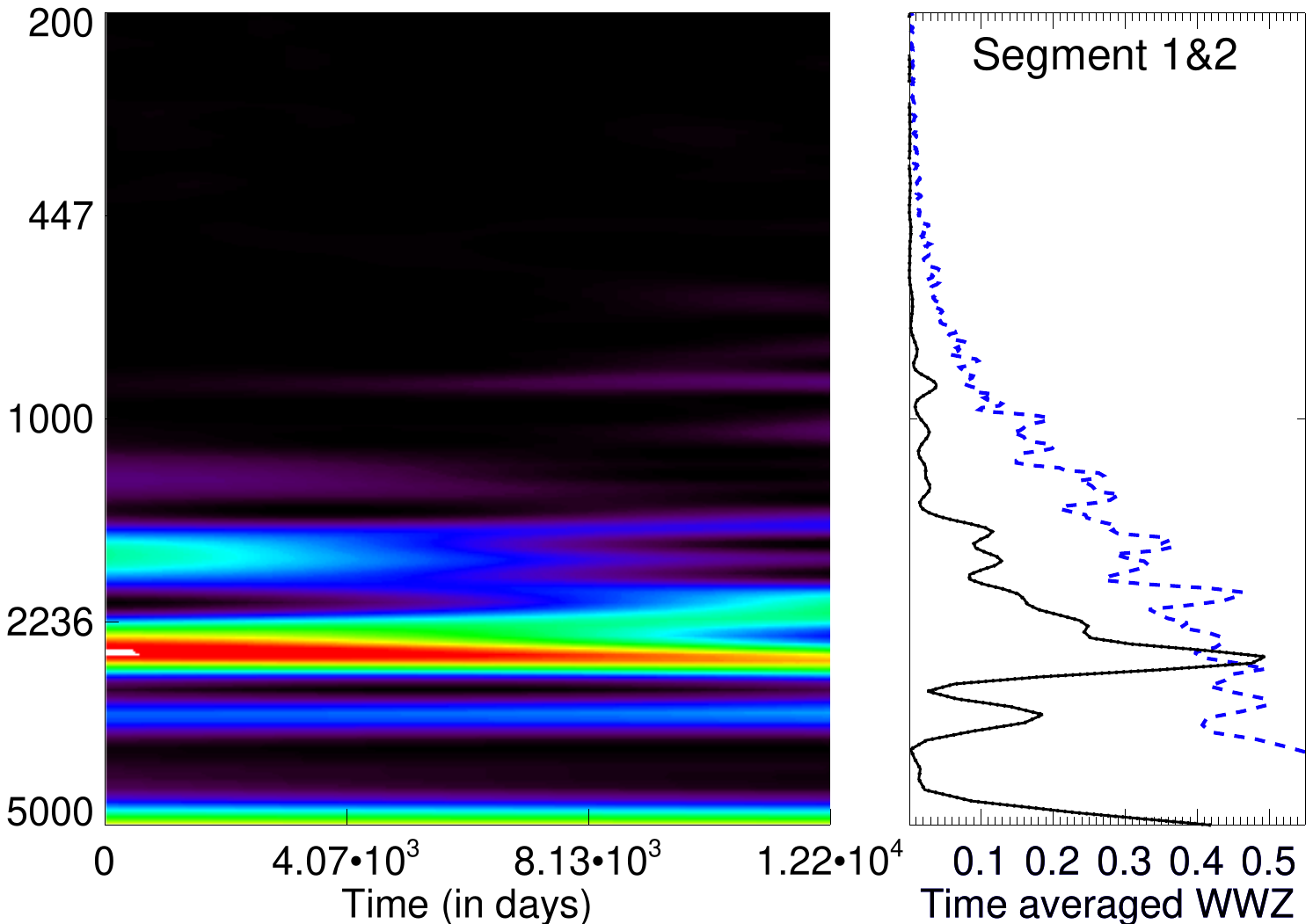}\includegraphics[angle=90, scale = 0.3]{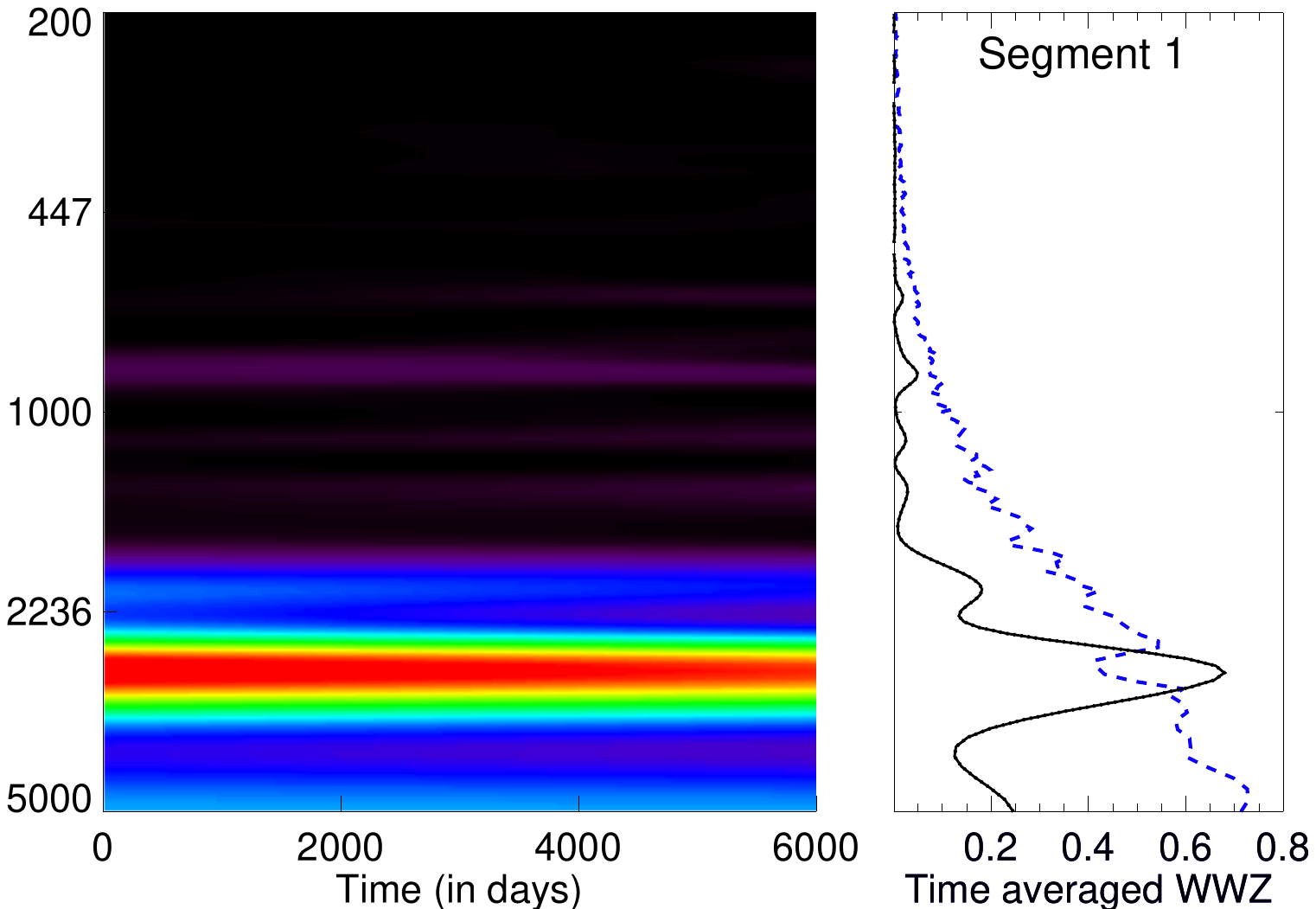}\\
\caption{As in Fig.\ 3 for the 8 GHz light curve.}
\end{figure*}\label{fig:8}

\begin{figure*}[t]
\centering
\includegraphics[scale=0.35]{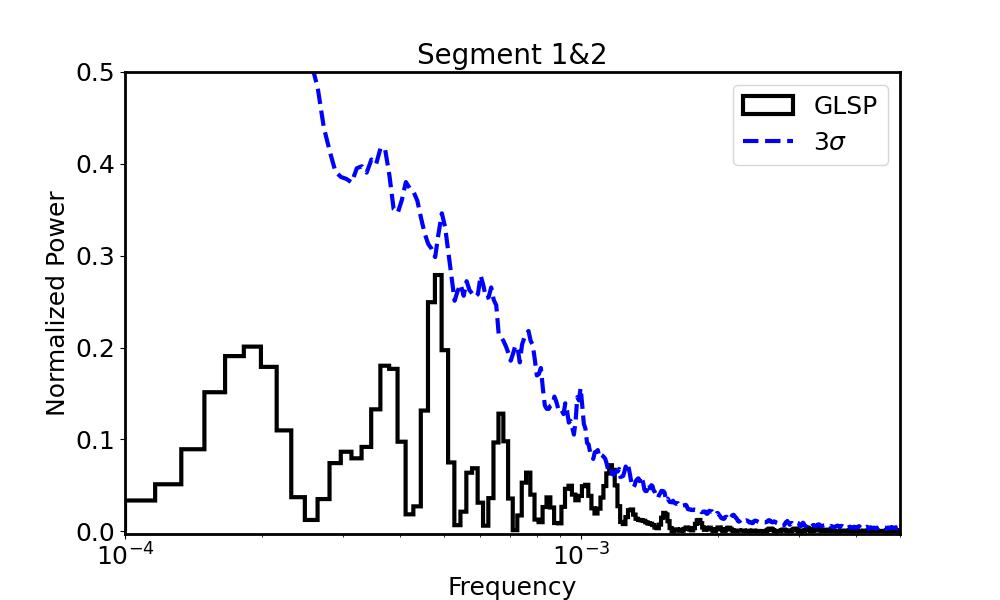}\includegraphics[scale=0.35]{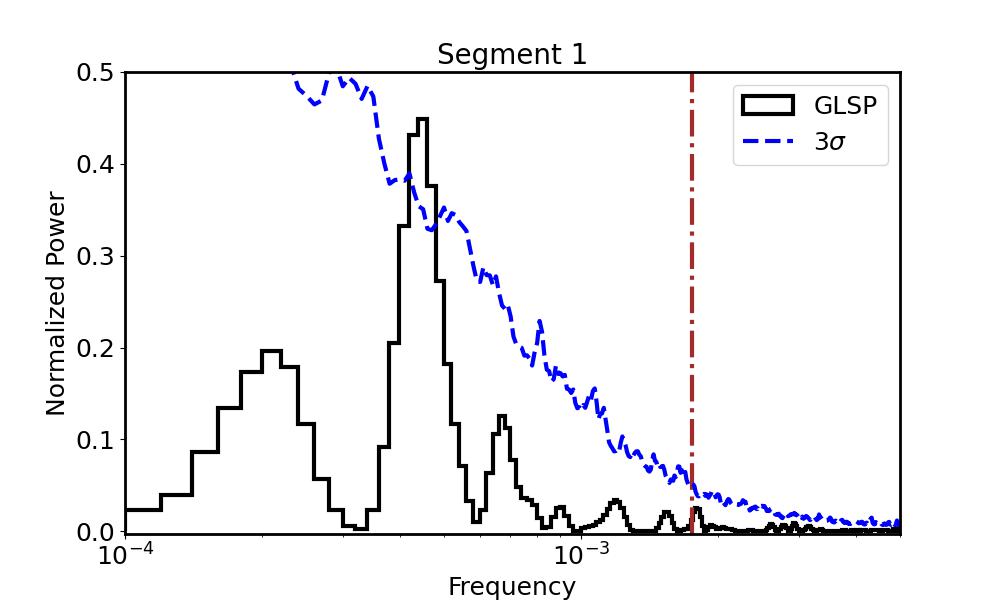}\\
\vspace{-1.0cm}
\includegraphics[angle=90, scale = 0.3]{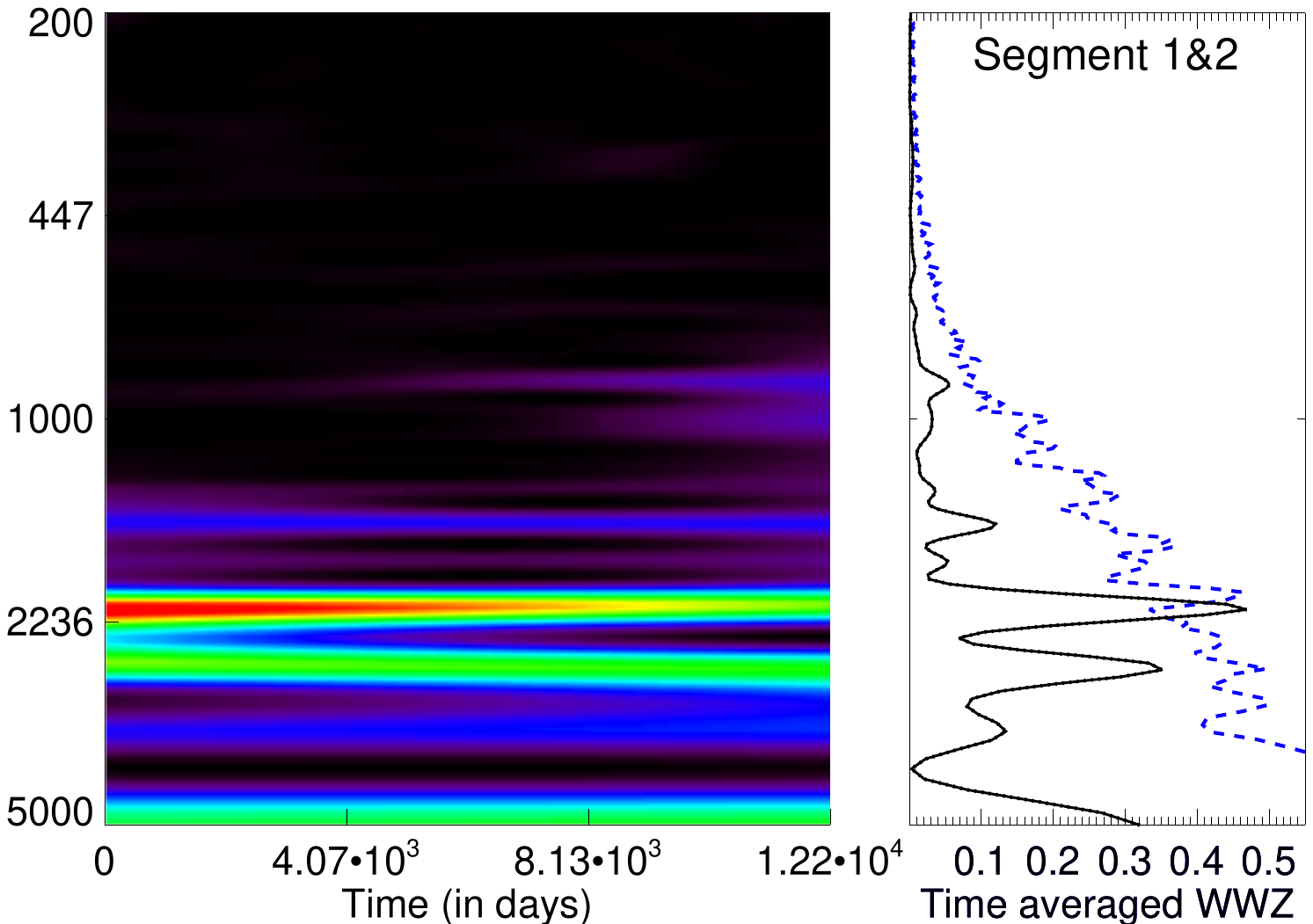}\includegraphics[angle=90, scale = 0.3]{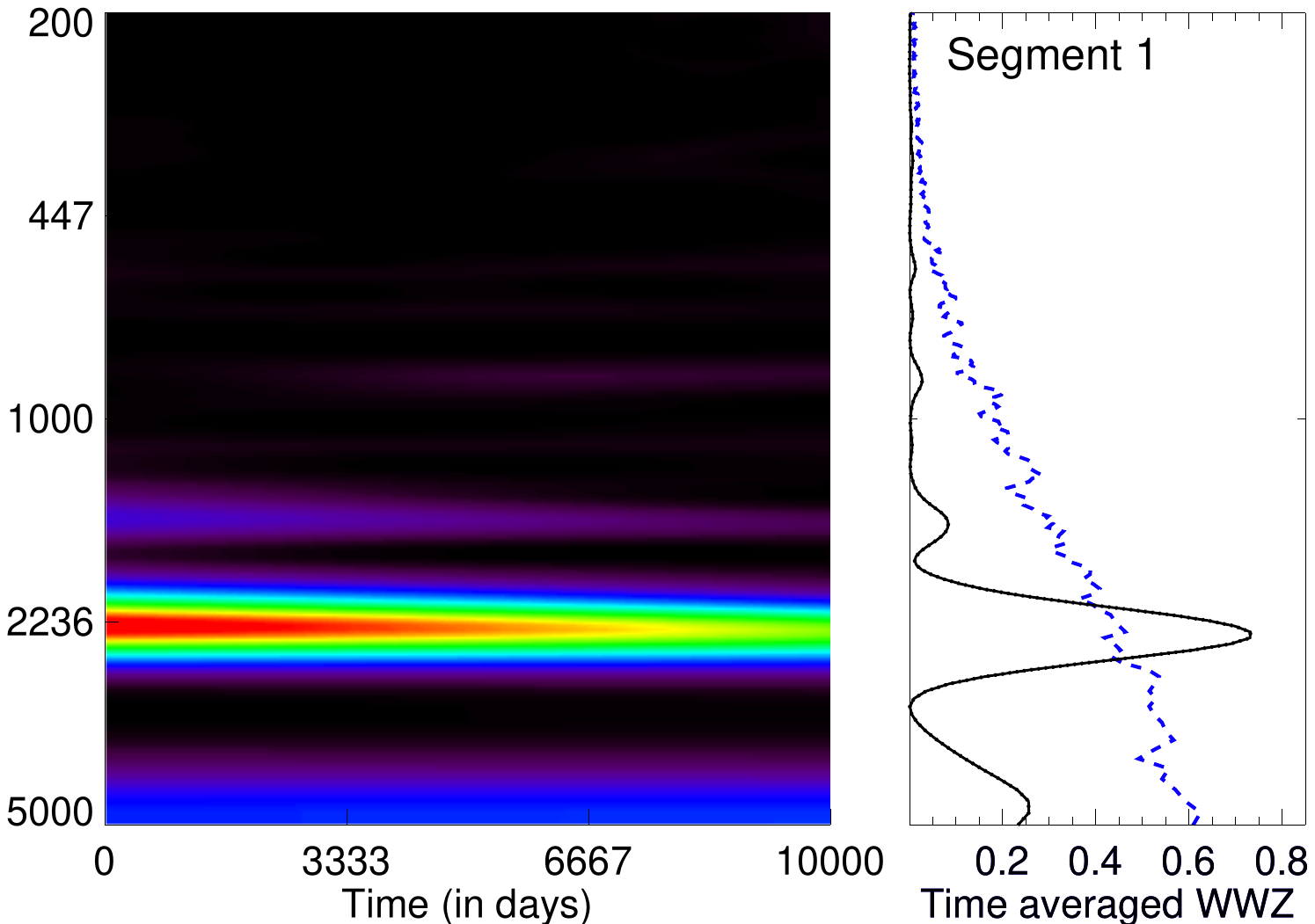}\\
\caption{As in Fig.\ 3 for the 14.5 GHz light curve. The dashed-dotted brown line in the GLSP plot of Segment 1 marks the possible QPO period at around 600 days.}
\end{figure*}\label{fig:14}

\section{Observations and Analysis Techniques}\label{sec:obs}
\noindent
The decades-long flux density observations at 4.8, 8.0, and 14.5 GHz are obtained from the University of Michigan Radio Astronomical Observatory \citep[UMRAO;][]{1985ApJS...59..513A}. The telescope at UMRAO is a 26-m equatorially mounted paraboloid and is equipped with radiometers operating at the aforementioned frequencies. Please see \cite{2021MNRAS.501.5997T} for more details about the observations from these three radio frequencies. The flux density observation at 4.8 GHz covers more than 30 years from 1978 to 2012. The radio observation at 8 GHz spans more than 45 years from 1966 to 2012. The 14.5 GHz UMRAO observations were initiated in 1974 and continued for almost 40 years till 2012. The 22-m radio telescope (RT-22) at Crimean Astrophysical Observatory \citep[CrAO;][]{2006ASPC..360..133V} was employed to monitor 3C 454.3 at 22.0 and 36.8 GHz for the period spanning from 1980 to 2013. 3C 454.3 is also observed at 37.0 GHz by a 14-m radio telescope operated by Aalto University Mets{\"a}hovi Radio Observatory in Finland. In order to produce a denser and longer data set, the observations taken at Mets{\"a}hovi and those from RT-22 at 36.8 GHz were integrated. For details about the data reduction and analysis of Mets{\"a}hovi data, please refer to \cite{1998A&AS..132..305T}. \\
\\
Fig.~1 shows the at least three-decade-long radio flux density observations at 4.8, 8.0, 14.5, 22.0, and 37.0 GHz. Upon visual inspection, the light curves display flux modulations which could be an indication of a  QPO. The flaring activity after 2007 can also be visually observed in light curves at all frequencies. The \textcolor{black}{relative} \black{amplitudes} of these flares increase from lower to higher frequency observations. In this work, we employ  generalized Lomb-Scargle periodogram (GLSP) and Weighted Wavelet Z-transform (WWZ) methods to confirm  plausible modulations in the light curves and to calculate their significance. We will briefly describe these methods in the following subsections. 

\begin{figure*}
\centering
\vspace{-0.1cm}
\includegraphics[scale=0.3]{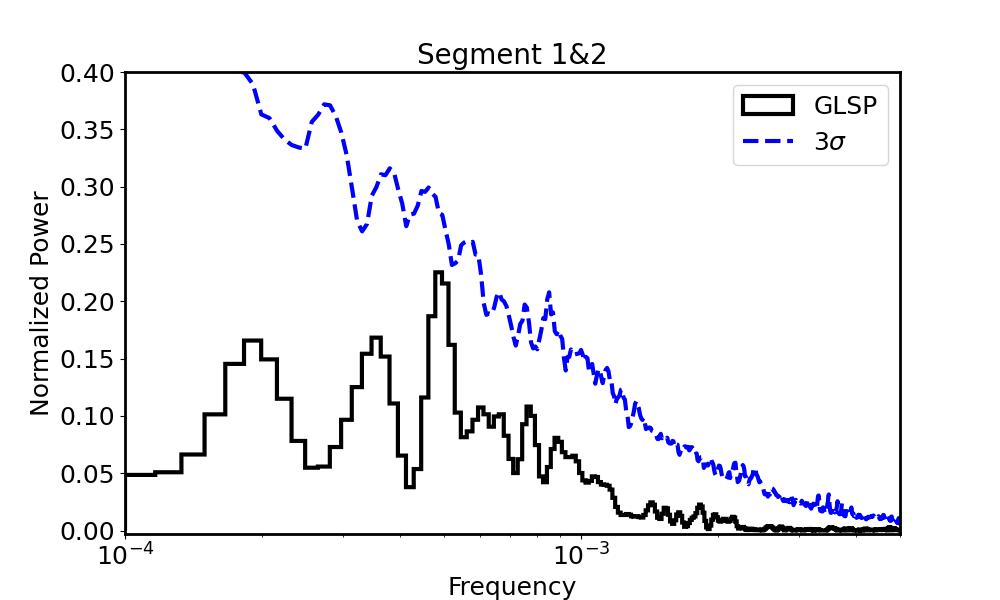}\includegraphics[scale=0.3]{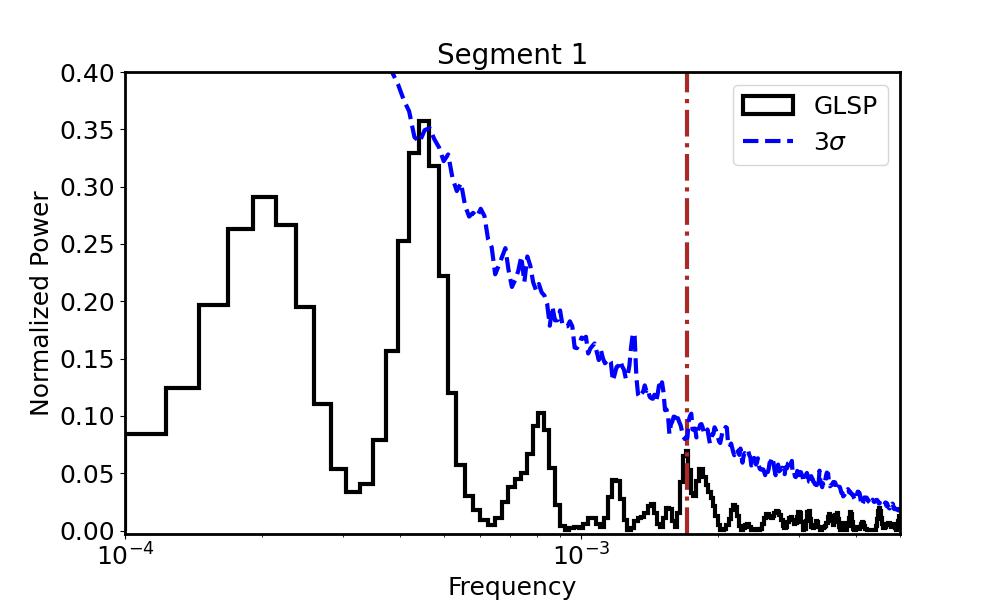}
\includegraphics[angle=90, scale = 0.3]{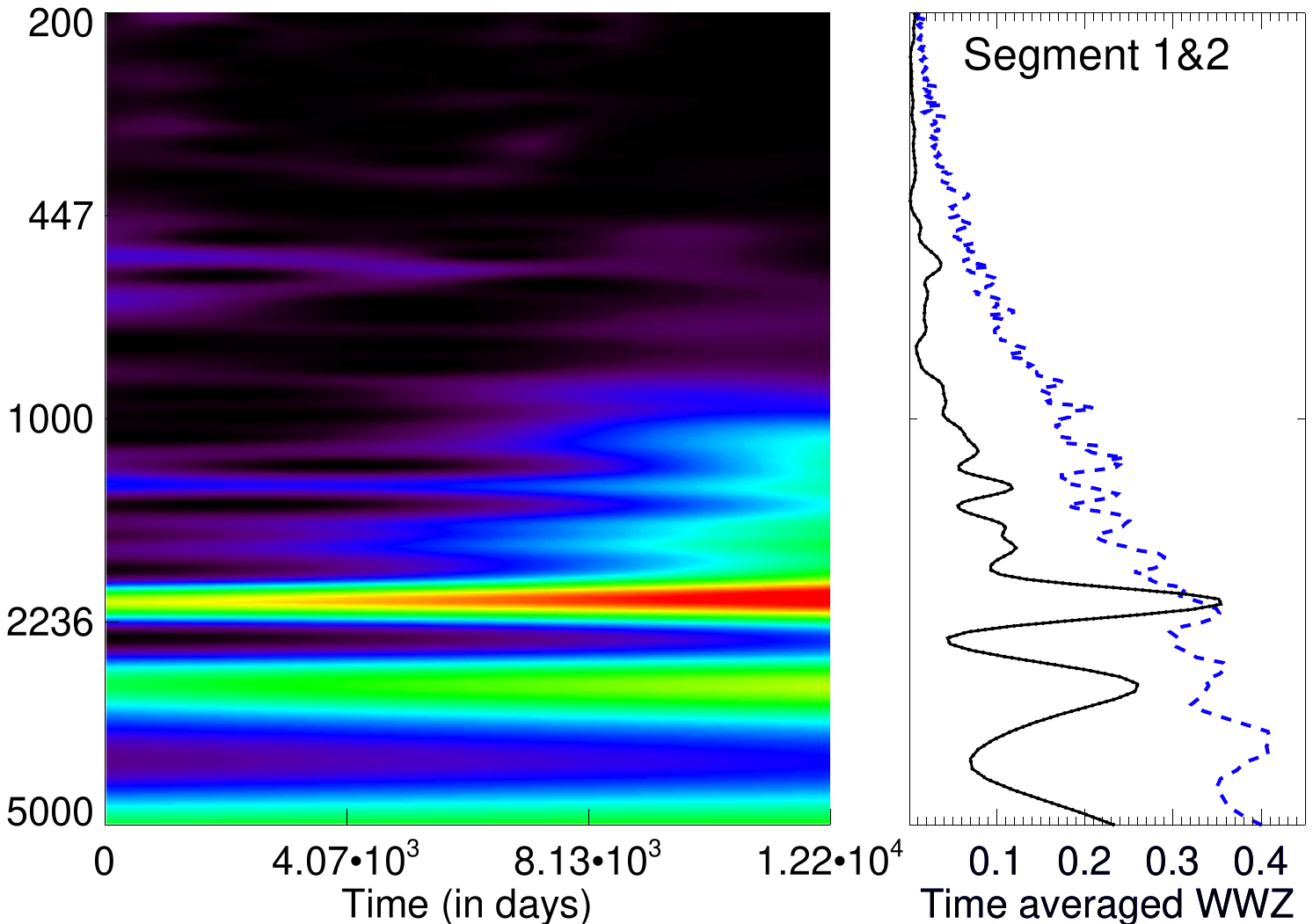}\includegraphics[angle=90, scale = 0.3]{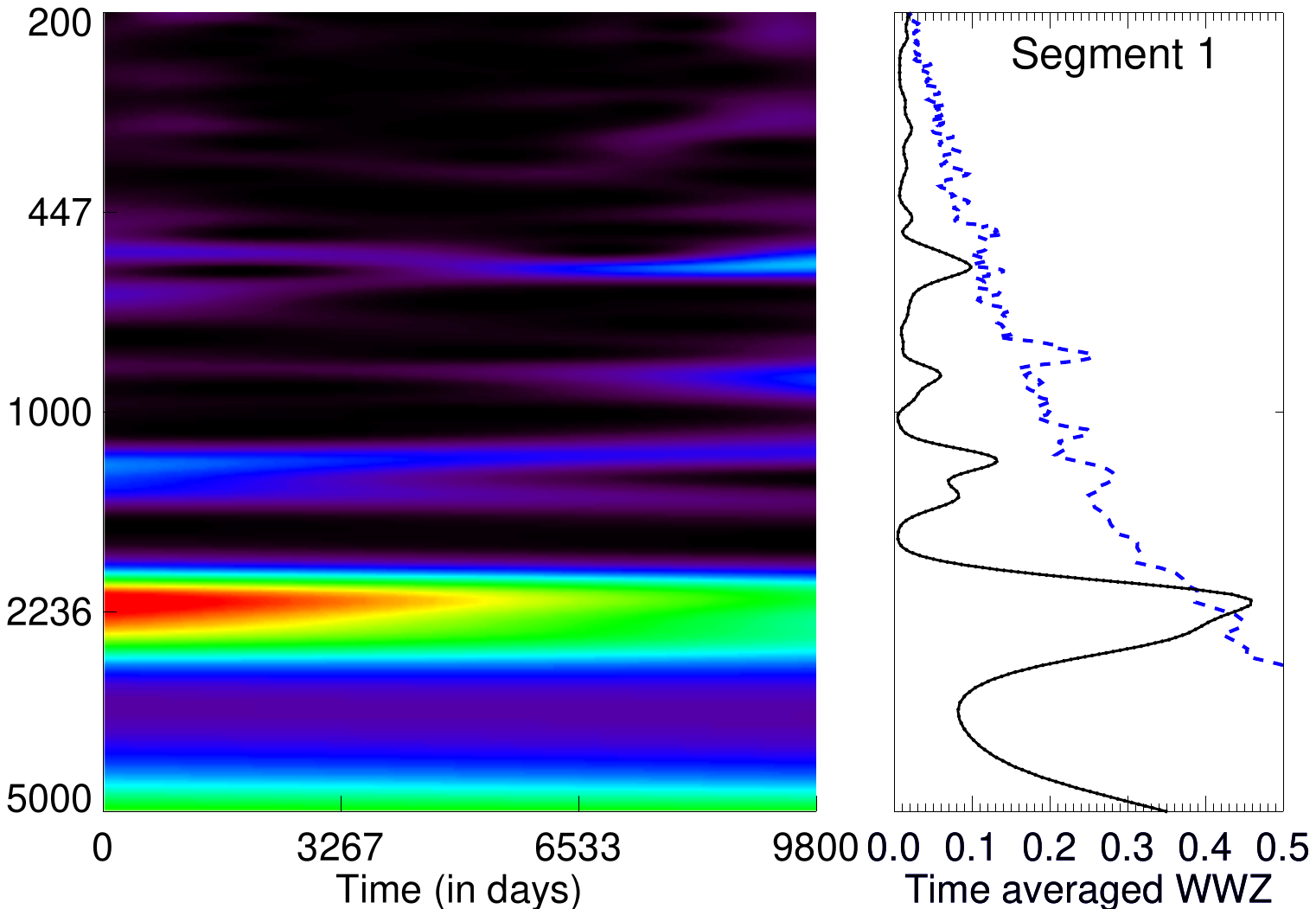}\\
\caption{As in Fig.\ 3 for the  22 GHz light curve. The dashed-dotted brown line in the GLSP plot of Segment 1 marks the possible QPO period at around 600 days.}
\end{figure*}\label{fig:22}

\subsection{Generalized Lomb-Scargle Periodogram}\label{sec:glsp}
\noindent
Periodograms are the classical technique for detecting any intrinsic periodic signal present in the data. The Lomb-Scargle Periodogram \citep[LSP;][]{1976Ap&SS..39..447L, 1982ApJ...263..835S} is a commonly used periodogram technique for unevenly sampled data. 
Here, we implement the generalized LSP (GLSP) routine from the {\tt PYSTRONOMY} package which is described in \citet{2009A&A...496..577Z}. See \citet{2024MNRAS.527.9132T} for details. 
We used the oversampling factor of 4.0 which means that the number of frequencies considered in this work is equal to twice the number of data points in the observation. We have also calculated the GLSP for other oversampling factors and found that the results are consistent within 3\%. \\
\\

\black{For significance calculations, the first step is to simulate the light curves using statistical properties similar to that of observations. Generally, only power spectral densities (PSDs) are used to simulate the light curves \citep[e.g.][]{1995A&A...300..707T}. These methods assume \black{that the probability density functions (PDFs) of the flux values are} normally distributed. However, if there is a ``burst-like" event during the observation, the PDF significantly deviates from the normal distribution, \black{becoming long-tailed, which therefore} limits the use of such methods for simulating light curves. \black{Hence,} using both PDF and PSDs are necessary to simulate the light curves \citep{2013MNRAS.433..907E}.} \\
\\
\black{Fig.\ 2 shows the \black{flux PDF} distributions for the light curves at the 5 radio frequencies analyzed in this work and their fits with normal and log-normal distribution. Lognormal \black{distributions are} commonly used to model AGN light curves \citep{2018ApJ...857..141S, 2020ApJ...891..120B}  \black{as the observations include ``flares" and display long-tailed distributions.} 
The flux distribution at 5 GHz is \black{essentially equally well fitted with normal and lognormal distributions but}  as we go to higher frequencies, the \black{the fluxes become substantially better fit by  lognormal distributions, indicating greater dominance by flare-like processes at higher radio frequencies. }}\\
\\ 
In this work, we used \textcolor{black}{broken power laws to model the power spectral density (PSD), $P(\nu)$, at frequency $\nu$ which is \black{thus} defined as }

\begin{equation}
\begin{split}
    P(\nu) &= N \left(\frac{\nu}{\nu_b}\right)^{-\alpha} + C,~~~ \nu > \nu_b;\\
           &= N \left(\frac{\nu}{\nu_b}\right)^{-\beta} + C,~~~ \nu < \nu_b~.
\end{split}
\end{equation}
\noindent \textcolor{black}{Here $N$ is the normalization, $\nu_b$ is \black{the break} frequency and $C$ is the instrumental white noise; $\alpha$ and $\beta$ are the  indices for the high-frequency part and low-frequency part, respectively.} \\
\\
\black{We simulate 10,000 light curves with similar statistical, flux distribution, and power spectral properties of an observation using the code\footnote{https://github.com/samconnolly/DELightcurveSimulation} which uses the method described in \cite{2013MNRAS.433..907E}.} Then, we calculate the GLSPs of all simulated light curves. 
The significance intervals are estimated using this normal power spectrum distribution at each frequency. See \citet{2024MNRAS.527.9132T} for details. 


\subsection{Weighted Wavelet Z-transform Analysis}
\noindent
The wavelet analysis technique is commonly used to study any QPO signals present in the data in both frequency and time domains. 
The Weighted Wavelet Z-transform (WWZ) is a wavelet technique that is well suited for real observations having sparse and uneven sampling \citep[e.g.][]{2005NPGeo..12..345W}. 
In this work, we used the publicly available WWZ software\footnote{\url {https://www.aavso.org/software-directory}} as employed recently in similar studies \citep[see][and references therein]{ 2017ApJ...847....7B, 2017ApJ...849....9Z,2021MNRAS.501.5997T}. See \citet{2024MNRAS.527.9132T} for details. \\
\\
If the WWZ power is marginalized over time, one gets the WWZ power in the frequency plane which is essentially a periodogram,
commonly referred to as the time-averaged WWZ, which usually follows a power law, similar to the Lomb-Scargle periodogram used in this work. To calculate the significance of any nominal  QPO signals, we followed the method described in \black{the previous subsection}. 

\subsection{Z-transformed Discreet Correlation Function}
\noindent

The cross-correlation function \citep{1988ApJ...333..646E} is a commonly used technique to study the emission mechanisms in AGNs by estimating the correlation between their variations in different bands. The z-transformed discrete correlation function (ZDCF) is an improved method  for computing the correlation function for unevenly sampled and irregular light curves \citep{2013arXiv1302.1508A}. In these ZDCF computations,  equal population binning and Fisher's z-transform method are applied to the DCF approach of \citet{1988ApJ...333..646E}. In this work, we employed a version of the ZDCF algorithm\footnote{https://www.weizmann.ac.il/particle/tal/research-activities/software} that has been implemented in recent studies of AGN \citep{2022Natur.609..265J, 2024MNRAS.52710168P}.

\section{Results}\label{sec:res}

\subsection{Confirmation of a $\sim$2000 d QPO}
\noindent
Fig.~1 shows the light curve of 3C 454.3 taken at 4.8 GHz in blue. The whole observation is divided into two segments. The second segment (after 2007) is  dominated by flare features that we suggest originate from a process different from the stochastic processes that occurred in the first segment of the observation. To assess the effect of these flares on the variability exhibited by this source, we calculate the power spectrum for two cases: the whole light curve (Segments 1 \& 2) and Segment 1 alone. These are plotted in the upper panel of Fig.~3. For the whole light curve,  a QPO signal at 0.00049 days$^{-1}$, corresponding to around 2040 days, is found to be at least 3$\sigma$
significant.  For Segment 1, a QPO signal at a similar frequency (0.00045 days$^{-1}$) is found at greater than  3$\sigma$ significance.
In the periodogram of the whole light curve, there is one more peak adjacent to the claimed one which is not present in the power spectrum of the first segment. This additional peak at lower frequencies appears to be contributed by the flares in Segment 2. \\  
\\
In the bottom panel of Fig.~3, WWZ power is plotted in the time-frequency plane for both the whole light curve and just for Segment 1. In each case, the left panel shows the color-color diagram of WWZ power. 
For the whole light curve, WWZ power is most concentrated at around 2080 days, and while it is persistent throughout the observation, it gets weaker during the last half of the entire span of the observations.  As in the GLSP, there is a weaker signal centered at the frequency of 0.00038 days$^{-1}$ which originates after the onset of the observations and continues until their conclusion. 
In Segment 1 the power is also concentrated around 2200 days, and this feature is more persistent in this segment than during the whole light curve, and no weak feature is present around the lower frequency. The right panel shows the 
WWZ time-marginalized periodogram which also indicates a significant oscillation at more than 3$\sigma$ confidence. \\
\\
The orange points in Fig.~1 correspond to the light curve collected at 8 GHz. The flare towards the end of the observation is even more significant than for the 4.8 GHz observation but the other flux variations are very  similar to those seen in the lower frequency observations. The upper panel of Fig.~4 shows the power spectrum analysis. For the whole light curve, a signal at the frequency of $\sim$~0.00048 days$^{-1}$ is found to exceed 3$\sigma$ significance using a simple power law. There are other peaks at higher frequencies but none are  significant. In the wavelet plot for the whole light curve, a signal exceeding 3$\sigma$ in the time-averaged WWZ is observed to be strong at the beginning of the observation but it weakens towards the end of the observations. We see that the strong signal appears to bifurcate around the middle of the observations and eventually becomes two weaker signals. This could be the result of the flaring activity at the end of observation. When only Segment 1 of the light curve is analyzed, the QPO peak at $\sim$~0.00048 days$^{-1}$ is stronger than that seen during the entire light curve, exceeding 3$\sigma$ regardless of the background model. In the wavelet plot, there is only one strong signal present, centered at 2380 days. \\
\\
Fig.~1 and Fig~5 show the 14.5 GHz light curve (green) and the GLSP and WWZ analysis results for the whole light curve and Segment 1, respectively. After early 2007, there is a clear indication of flaring activity which has significantly higher flux relative to the earlier fluctuations and is also stronger compared to both the 4.8 and 8 GHz light curves. In the GLSP result for the whole light curve, the signal around 2125 days ($\sim$~0.00047 days$^{-1}$) \black{is marginally consistent with} 3$\sigma$ significance. In the GLSP plot of Segment 1, there is only one strong peak with at least 3$\sigma$ significance, which is slightly shifted to  $\sim$~0.00042 days$^{-1}$. The wavelet plots show similar behaviors to those of the 8 GHz light curve. The wavelet color-color diagram of the whole light curve shows a strong peak in the beginning which fades as it approaches the end of the observations. There is also a second peak of lesser significance.  The WWZ plot of Segment 1 shows a strong signal around 2325 days which persists throughout this extensive portion of the observations, though it does weaken with time. \\ 
\\
The 22 GHz radio light curve of 3C 454.3  is shown in Fig.~1 in red. At this frequency, the extended flare in Segment 2 is even more dominant. The fluxes in Segment 1 are similar to those at the lower radio frequencies, but they are significantly higher in Segment 2. 
The influence of these flares is easily seen in the periodogram plotted in the upper panel of Fig.~6. In the PSD of the whole light curve (left figure), the signal at the period of around 2080 days is found \black{with the marginal} 3$\sigma$ significance. 
If this baseline is affected by some other processes, then it is difficult to determine the appropriate significance levels, as is shown later in Section~\ref{cav} through simulations.  In the case of segment 1 (right figure), the peak at the frequency of 0.00048 days$^{-1}$ (2200 days) is found to have at least 3 $\sigma$ significance. \\  
\\
In the wavelet plot for the whole light curve (bottom panel), the WWZ power is mostly concentrated at the frequency of $\approx$0.00048 days$^{-1}$, similar to what is found in the GLSP analysis, with more than 3$\sigma$ significance as indicated in the time-averaged WWZ plot. For Segment 1, most power is concentrated around 2200 days, as also found in the GLSP method. One difference from the lower frequency data is that the WWZ for the entire  light curve has more power concentrated toward the end of the observation while the opposite is the case for Segment 1. \\ 
\\
At 37 GHz, the very powerful late flare is the most evident feature in the light curve which is shown in  Fig.~1 in violet. The count rate for Segment 2 is almost double  that of Segment 1; however, the fluctuations seen at lower radio frequencies are still evident in Segment 1. In the upper panel of Fig.~7, the periodogram is plotted for the whole light curve and for Segment 1. Similar to 22 GHz, the ``red noise" level for the whole light curve is unstable, which results in less confident determinations of the underlying power law models. So, no peak in its periodogram is found that has 3$\sigma$ significance.
In the wavelet color density diagram,  the greatest power is concentrated toward the end of observation at multiple periods, more than that for 22 GHz. Although the signal at 0.00046 days$^{-1}$ has 3$\sigma$ significance (as seen in the time-averaged WWZ plot), it is strong only in the first half of the observation after which it starts to weaken and then more power around that frequency can be seen at the end of the observation. 
For Segment 1, the signal at 0.00042 days$^{-1}$ is persistent throughout the observation and it exceeds 3$\sigma$. No additional signals at the end of the observation are seen in this plot.\\
\\
\black{Table 1 lists the best-fit parameters for the broken power laws used to fit the \black{the power spectra of the observations.} 
For all the radio frequencies, the high-frequency  index $\alpha$ is \black{as steep or steeper} than the typical value found for blazars when analyzing Segments 1 \black{and 2 together, ranging between 1.9 and 4.5. However, when only Segment 1 is considered, for all radio bands, $\alpha$ is found to have values in the range 2--3 , within errors.  The values of the slopes of the lower-frequency portion of the PSD, $\beta$, are all between 0.4 and 1.2.}  } 

\begin{figure*}
\centering
\includegraphics[scale=0.3]{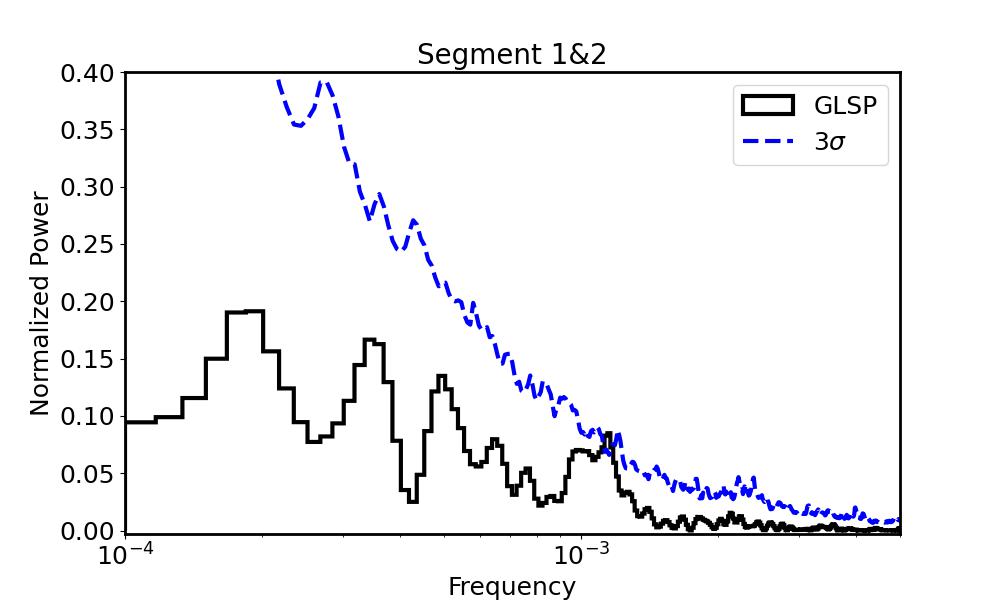}\includegraphics[scale = 0.3]{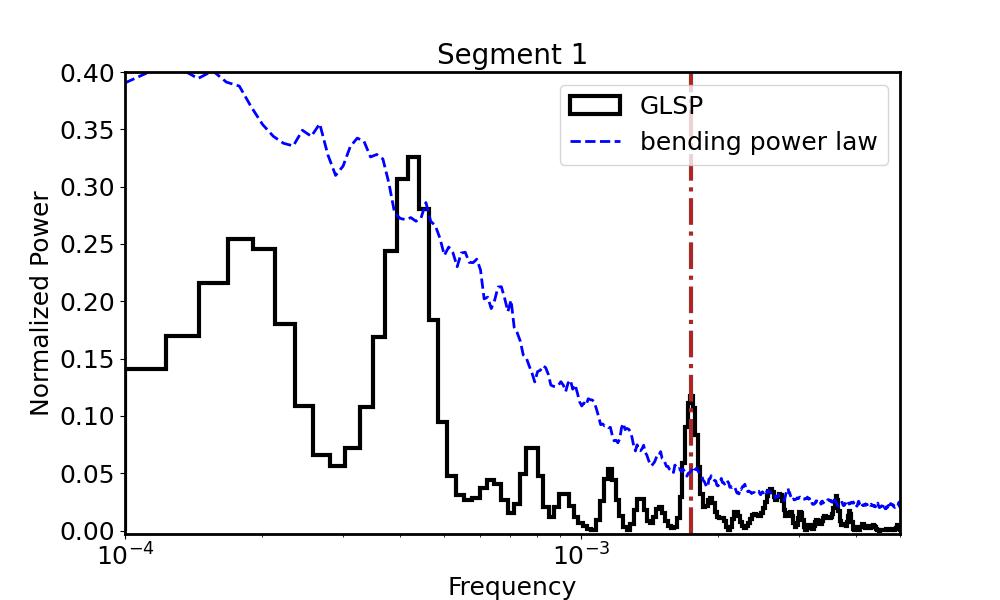}
\vspace{-1.8cm}
\includegraphics[angle=90, scale = 0.3]{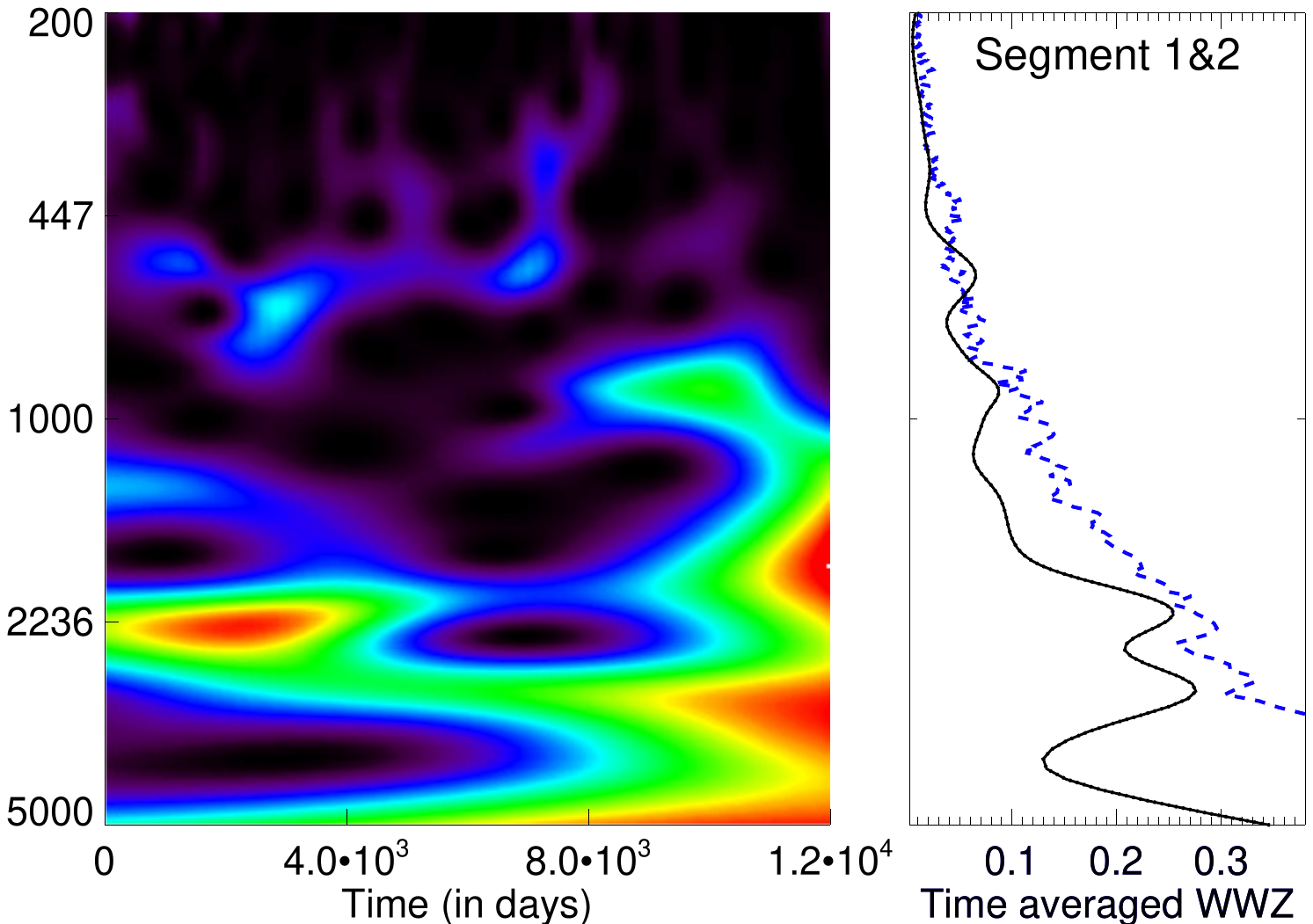}\includegraphics[angle=90, scale = 0.3]{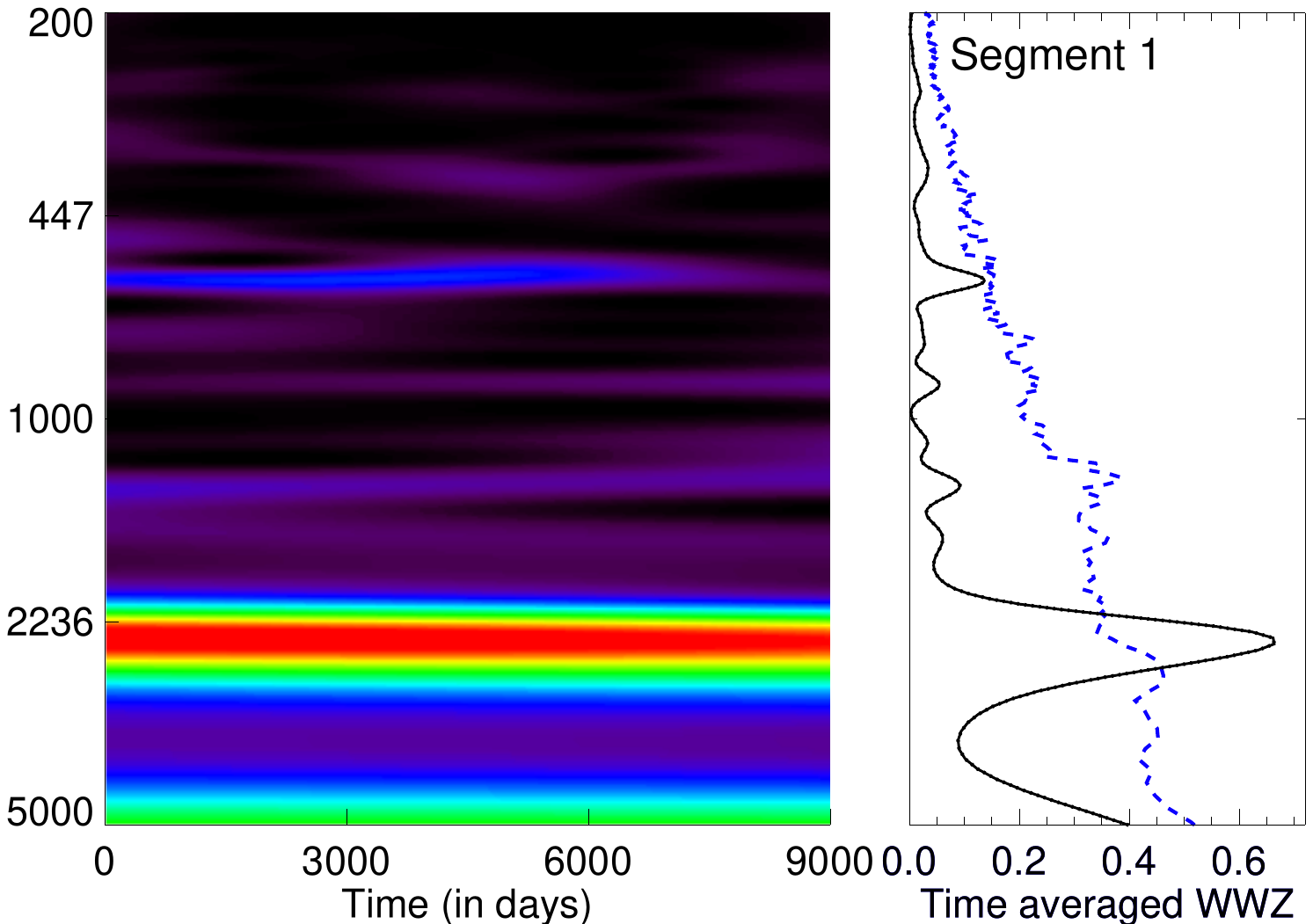}
\caption{As in Fig.\ 3 for the 37 GHz light curve. The \black{brown dot-dashed line in the GLSP plot of Segment 1} marks the likely QPO at $\sim$600 days.}
\end{figure*}\label{fig:37}

\begin{table*}
 \caption{Best-fit parameters for the broken power used to fit the light curves analyzed in this work. }
\begin{tabular}{|c|c|c|c|c|c|c|}
\hline
Frequency  & Segment&$\nu_{b}$ ($\times 10^{-4}$) & $\beta$ & $\alpha$  & N & C \\\hline
\multirow{ 2}{*}{4.8}& 1\&2 &0.66$\pm$0.08&0.83$\pm$0.09&1.87$\pm$0.21 &0.032$\pm$0.002&0.011$\pm$0.001\\
& 1& 0.1$\pm$0.01&1.21$\pm$0.14&1.73$\pm$0.21&0.037$\pm$0.04&0.011$\pm$0.001\\\hline
\multirow{ 2}{*}{8}& 1\&2 &2.82$\pm$0.23&0.81$\pm$0.1&4.52$\pm$0.32 &0.08$\pm$0.006&0.063$\pm$0.005\\
& 1& 1.62$\pm$0.13&0.97$\pm$0.08&3.4$\pm$0.43&0.078$\pm$0.09&0.033$\pm$0.004\\\hline
\multirow{ 2}{*}{14}& 1\&2 &3.3$\pm$0.34&0.72$\pm$0.08&4.02$\pm$0.39 &0.48$\pm$0.054&0.066$\pm$0.006\\
& 1& 2.21$\pm$0.24&0.52$\pm$0.06&2.83$\pm$0.31&0.99$\pm$0.11&0.034$\pm$0.003\\\hline
\multirow{ 2}{*}{22}& 1\&2 &3.9$\pm$0.38&0.59$\pm$0.06&4.23$\pm$0.23 &2.37$\pm$0.22&0.406$\pm$0.008\\
& 1& 4.9$\pm$0.05&1.21$\pm$0.13&2.08$\pm$0.20&0.002$\pm$0.001&0.070$\pm$0.007\\\hline
\multirow{ 2}{*}{37}& 1\&2 &2.03$\pm$0.19&1.24$\pm$0.11&2.61$\pm$0.25 &0.015$\pm$0.002&0.316$\pm$0.025\\
& 1& 4.4$\pm$0.04&0.35$\pm$0.03&2.51$\pm$0.28&3.093$\pm$0.03&0.292$\pm$0.034\\\hline

\end{tabular}
\end{table*}

\subsection{A $\approx$600-day signal}
\noindent
Interestingly, a signal at around 600 days is also detected in these radio observations, but it is only significant in the light curves measured at higher radio frequencies. This apparent QPO is also suppressed by the flaring processes that began after early 2007. At 4.8 GHz, a peak at around 600 days is found in the GLSP result of the combined Segments 1 and 2 but it is not significant. This result also holds when only Segment 1 is analyzed. As this feature was not detected in the WWZ analysis, we do not claim it is present at this radio frequency. The same result also holds for the \black{8 GHz radio light curve}. \\
\\
\black{\black{A possible} signal, although with significance less than 3$\sigma$, is observed to emerge around 600 days in the Segment 1 light curve at 14 GHz, which is marked with \black{a brown dashed line in the GLSP plot in Fig.~5.} For 22 GHz, no significant signal \black{around that period} is observed when the entire observation is analyzed. However,  when only Segment 1 is examined, the significance is nearly 3$\sigma$ \black{(see Fig.~6), which is higher} than that obtained for 14 GHz. }
While this signal is rather weak in the WWZ analysis spanning Segments 1 and 2, it is clearly seen in the wavelet \black{plot in Fig.~6} for Segment 1. It persists throughout this portion of the observations and becomes stronger toward the end of it. The time-averaged WWZ signal reaches a 3$\sigma$ significance.  We conclude this feature was suppressed by the flaring  which became dominant after 2007.\\ 
\\
At 37 GHz, this feature at around 600 days is absent in the periodogram of the full set of observations. Instead, a signal around 870 days  is present and appears \black{to be roughly 3$\sigma$} significant. The $\approx$~600-day signal can be observed in the wavelet plot for all the data, albeit below 3$\sigma$ significance. \black{However, the 600 day signal is seen in the GLSP plot  of Segment 1 in Fig.~7} with more than 3$\sigma$ significance, which actually makes this signal the most significant one at this radio frequency. The wavelet plot of Segment 1 also shows this signal with a significance exceeding 3$\sigma$. This 600-day signal corresponds to around 15 cycles of observations in the data up to 2007, indicating it is \black{a rather strong} candidate QPO period, at least in the light curves at 22 and 37 GHz.

\subsection{ZDCF}
\noindent
Fig.\ 8 shows the ZDCF analysis of the light curves used in this work for Segment 1 (top) and Segment 2 (bottom).  The 4.8 GHz radio light curve is taken as the base light curve with respect to which the cross-correlation is calculated. In Segment 1, the auto-correlation of the 4.8 GHz radio light curve with itself also illustrates the quasi-periodic patterns discussed above with the first peak (at non-zero lag) at around 2400 days, and a ZDCF value of 0.25. The maximum ZDCF values of the 4.8 \& 8 GHz and 4.8 \& 14.5 GHz cross-correlations are 0.92 and 0.85, respectively, with negative lags, indicating that the 4.8 GHz light curve lags behind those at the higher frequencies, by 331 d and 472 d, respectively. The peak ZDCF values against 4.8 GHz are found to be smaller at 22 GHz (0.69) and 37 GHz (0.47), with respective negative lags of 1010 d and 1162 d.  The lower peak ZDCF values at 22.0 and 37.0 GHz \black{could} arise from the lower fluxes at those higher frequencies, along with the confounding effects of the $\sim 600$d QPO \black{apparently present in} them. \textcolor{black}{Another possible reason for such behavior could be differences in the sizes of emission regions, which can be expected to be smaller at higher radio frequencies as well as the difference in opacity at different radio frequencies. This behavior is consistent with}  the classic van der Laan adiabatically-expanding source model \citep{1966Natur.211.1131V}. \black{More plausibly, if the variable radio emission arises from instabilities that weaken as they propagate downstream in the jet, the lower frequency emission, which arises further downstream would be both delayed and reduced in amplitude.}\\ 
\\
In Segment 2, the maximum ZDCF value is very high for all radio frequencies, as might be expected when only one major and one minor flare are present in that interval.  The lags at 8 and 14.5 GHz are nearly the same as in Segment 1, but those at 22 and 37 GHz are smaller, supporting the hypothesis that a different physical mechanism is responsible for the flares in Segment 2, which are strongest at the highest frequencies. \textcolor{black}{It is also possible that the same physical mechanism produces these flares but the  location or physical parameters producing the synchrotron emission are different. }

This increased flux in Segment 2 at high radio frequencies also could lead to masking of the probable $\sim$600d quasi-period which could only be recovered when analyzing Segment 1 individually.

\subsection{Caveats}\label{cav}
\noindent
In principle, the physical mechanism responsible for flaring could be different from the \black{usual stochastic processes occurring in jets or accretion disks of blazars that produce the run-of-the-mill variability.  If this is the case it would not be modeled appropriately with a (broken) power-law, as is} done for stochastic processes. Also, the statistical properties of the light curve are different during flare and non-flare periods \citep{2015MNRAS.452.2004M}. \black{In the previous section for estimating the significance of the claimed QPO signals found in the whole light curve (Segment 1\&2), we fit the periodogram of segment 1 \&2 jointly by a broken power law and then simulate the light curves. In this section, we want to assess how the significance estimates are affected if we model segment 1 and segment 2 separately.} To assess the effects of these issues on our results, we simulate the light curves in two pieces. In this work, we consider the whole 37 GHz radio light curve as the input observation. The first \black{is comprised of} the stochastic process model with a broken power law obtained by fitting the light curve of Segment 1 at 37 GHz. The second component \black{considered} the flares modeled in the same way as for the first component but obtained by fitting Segment 2. \black{ Modeling the flare itself is outside the scope of this work so, we chose the broken power-law for consistency.} Then we combine these two segments \textcolor{black}{in time} to form a single light curve. In this way, we simulate 10,000 light curves and follow the procedure described in the \textcolor{black}{Sec.~\ref{sec:glsp}} to obtain the desired confidence regime.\\
\\
We compare our results with the case where both segment 1 and segment 2 are fitted jointly with the underlying broken-power law model. 
\blue{Fig.~9}  shows 1$\sigma$ confidence regions for the power spectrum density obtained by simulating the light curves \black{fitting segment 1 and segment separately} which is denoted by the red curve. The blue region corresponds to the error region for the PSD obtained by fitting the light curve with a single red noise model throughout the observation. The error estimates for both cases are consistent with each other in the  frequency range of 0.0003 days$^{-1}$--0.0015 days$^{-1}$. As the claimed QPOs are found at frequencies higher than 0.0004 days$^{-1}$, this will have minimal impact on our calculations. At frequencies lower than 0.0003 days$^{-1}$, the confidence regions are significantly different.  The error in the case that includes flares is smaller compared to the case that excludes flares and is consistent with each other above a frequency of 0.0003 days$^{-1}$. Thus, fitting flares as a different stochastic process in the significance estimation would overestimate the errors for \black{temporal frequencies less than 0.0003 days$^{-1}$ but should not} affect the QPO signal analyzed in this work.

\section{Continuous Auto-regressive moving average (CARMA) analyses}
 \black{\black{To go beyond the basic modeling of PSDs in terms of a broken power-law, we have also performed CARMA analyses on these radio light curves.}  This is a method to model the light curves directly in the time domain and hence is not affected by the spectral distortions as in the case of frequency-domain analyses} \citep[e.g.][]{2014ApJ...788...33K, 2017MNRAS.470.3027K}. \black{A CARMA($p,q$) process is the solution of the set of stochastic differential equations where $p$ and $q$ respectively define the order of auto-regression and moving-average processes.} For a CARMA process to be stationary, it is necessary that $q<p$.
  \black{The majority of the long-term variability of AGNs is thought to originate in the accretion disk of the system \black{and in the outflows and jets which interact} with the surrounding matter and thus there is a \black{substantial possibility that the observed} variability becomes complex and non-linear in nature. Hence fitting the observations with differential non-linear equations of nth-order \black{should provide a better mode for such} non-linear processes.} \\
\\
 \black{We modeled the light curves analyzed in this work using CARMA($p,q$) \black{as implemented in the} {\tt Eztao} python package \citep{2022ApJ...936..132Y}, where $0\le p \le 7$ and $0\le q< p$. We fit each light curve with CARMA models with different values of $p$ and $q$ and then select the model for which the Akaike information criterion (AIC) \black{is minimized}. We then use that CARMA model to construct the periodogram and compare it with the Lomb-Scargle for a reality check. Fig. 10 shows the PSD constructed using the best-fit CARMA model (written in the plot) and also the binned Lomb-Scargle periodogram. The CARMA \black{derived PSDs all show their highest normalized powers} around 0.00038 days$^{-1}$  for all five radio frequencies which is consistent with the QPO frequency claimed in this work.}

\begin{figure*}[t]
\centering

\includegraphics[scale=0.6]{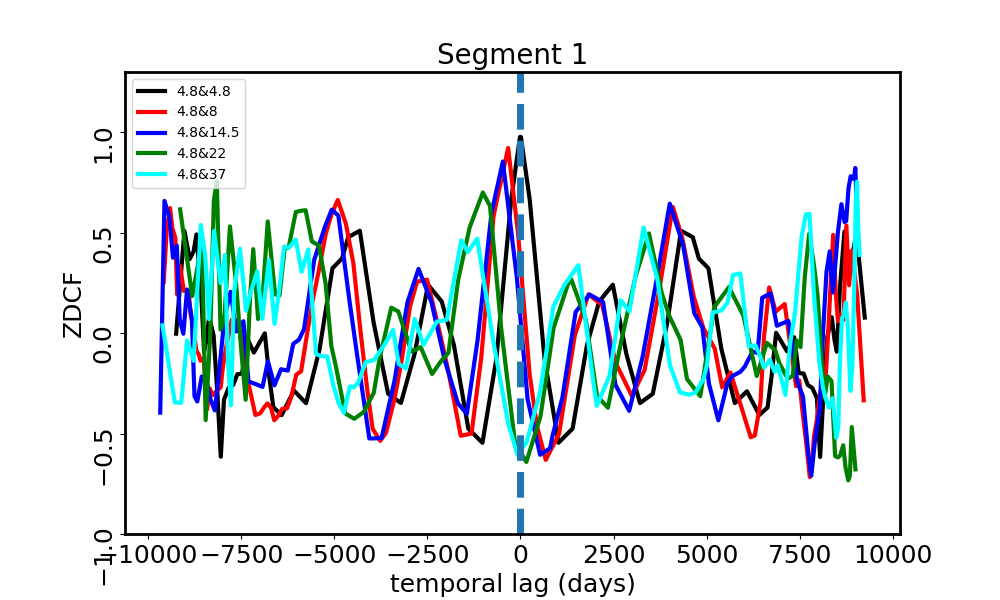}\\
\includegraphics[scale=0.6]{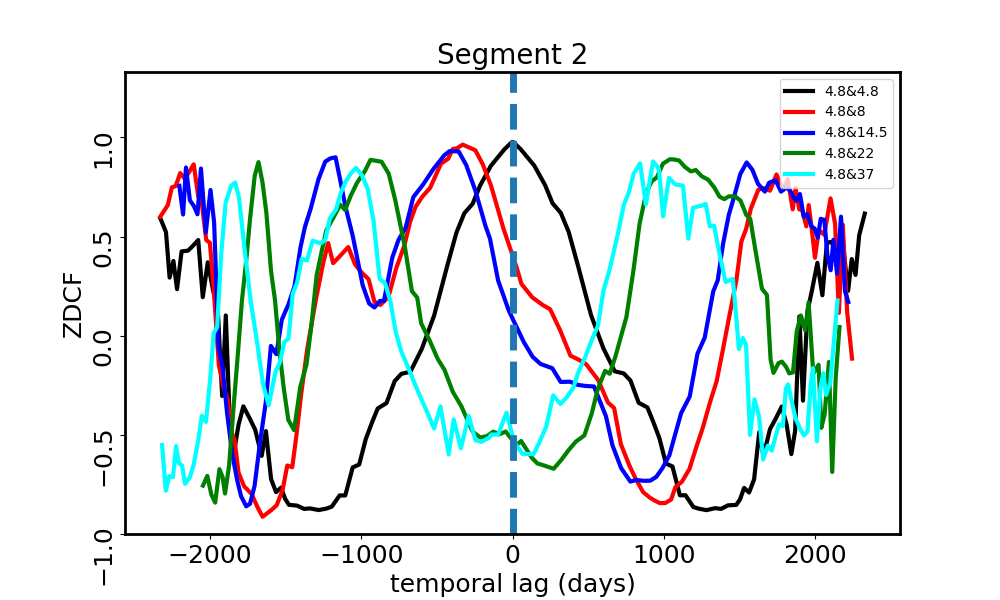}

\caption{Cross-correlations between the radio light curves in Segment 1 (pre 2007) and Segment 2 (post 2007). }
\end{figure*}\label{fig:ccf}

\begin{table*}
 \centering
 \caption{Likely QPO periods for different radio frequencies measured by different methods }
\begin{tabular}{|c|c|c|c|}
\hline
Frequency  & method & Segment 1 \& 2  & Segment 1 \\\hline

 4.8 GHz & GLSP & $2093\pm 52$& $2272\pm 62$  \\
 & WWZ & $2083\pm 26$ & $2223\pm 52$ \\ \hline
 8 GHz & GLSP & $2089\pm 30$& $2367\pm 70$  \\
 & WWZ & $2127\pm 55$ & $2380\pm 60$ \\ \hline
 14.5 GHz & GLSP & $2125\pm 27$& $2283\pm 56$  \\
 & WWZ & $2127\pm 63$ & $2325\pm 65$ \\ \hline
 22 GHz & GLSP & $2085\pm 34$ & $2260\pm 52$ \& \textbf{$580\pm 34$}  \\
 & WWZ & $2083\pm 26$ & $2127\pm 55$ \\ \hline
 37 GHz & GLSP & $2043\pm 27$& $2426\pm 64$ \& \bf{$582\pm 32$}  \\
 & WWZ & $2174\pm 64$ & $2439\pm 121$ \\ \hline
\end{tabular}
\end{table*}


\section{Discussion and Conclusions}\label{sec:dis}
\noindent
In this work, we analyzed $\approx$ 35 year-long observations of a blazar 3C 454.3 taken at the radio frequencies of 4.8, 8.0, 14.5, 22.0, and 37.0 GHz where the flux density at 4.8, 8.0, and 14.5 GHz are obtained from UMRAO and that at 22.0 and 37.0 GHz are from RT-22 and Aalto University Mets{\"a}hovi Radio Observatory. The possible periodic modulations in flux can be seen in the light curves as shown in \blue{Fig.~1}. GLSP and WWZ methods are used to assess the periodicity observed in these light curves. 
To calculate the desired significance level, the power spectrum at lower frequencies is modeled with red noise, which means the power is inversely proportional to the frequency raised to some power. We used \black{a broken power-law} to model the underlying red-noise stochastic process. Before 2007, the variability follows a stochastic process with an overlying QPO. After 2007, the light curves are dominated by a strong flaring process that is more significant at higher radio frequencies. We analyzed the observations with and without including this flaring period. \\
\\
A period of around 2000 days is detected at all five radio frequencies, irrespective of the inclusion of the later flaring period in our analysis. Table 1 lists the probable QPO periods found using GLSP and WWZ analysis methods for these radio frequencies. The detection is \black{considered strong if it is of at least \black{3$\sigma$ significance} in both the GLSP and  the time-averaged WWZ \black{analyses}} using \black{a broken power law} as the underlying red noise model. For full light curves, this period corresponds to $\approx$ 6 putative cycles with 12000 days as the temporal baseline of these observations. The strength of this signal increases at all the radio frequencies when the strong flaring period of Segment 2 is not included in the analysis, which suggests that \black{this additional flaring activity makes a large change to the  stochastic process and destroys or swamps any QPO}. This effect can also be detected in the WWZ color diagram where most of the power is concentrated at the end of the observation when including the flaring period and the diagram becomes chaotic as one examines the higher radio frequency light curves.\\
\\
Interestingly, a quasi-period of $\sim$~600 days is also observed with at least 3$\sigma$ significance at the frequencies of 22.0 and 37.0 GHz when the flaring period is excluded from the analysis. This period is also present in observation at lower radio frequencies but with less significance. This period corresponds to $\approx$ 16 putative cycles of observations. As the claimed quasi-periodic frequencies are more than 0.0003 days$^{-1}$, the error estimates on the periodogram are not affected by the flaring processes significantly
which was shown through simulations discussed in Sec.\ 3.4. \\

\begin{figure*}[t]
\centering

\includegraphics[scale=0.5]{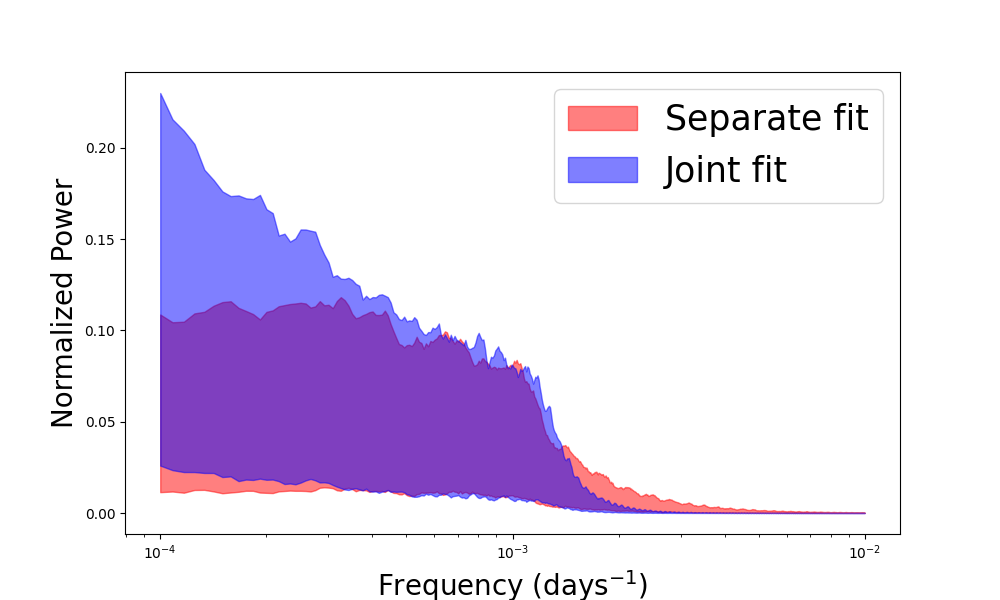}

\caption{Power spectrum simulations illustrating the effect of flaring activity on the error estimation using \black{broken power law} as underlying red noise models.  }
\end{figure*}\label{fig:flare}

\begin{figure*}[t]

\hspace{-1.2cm}\includegraphics[scale=0.28]{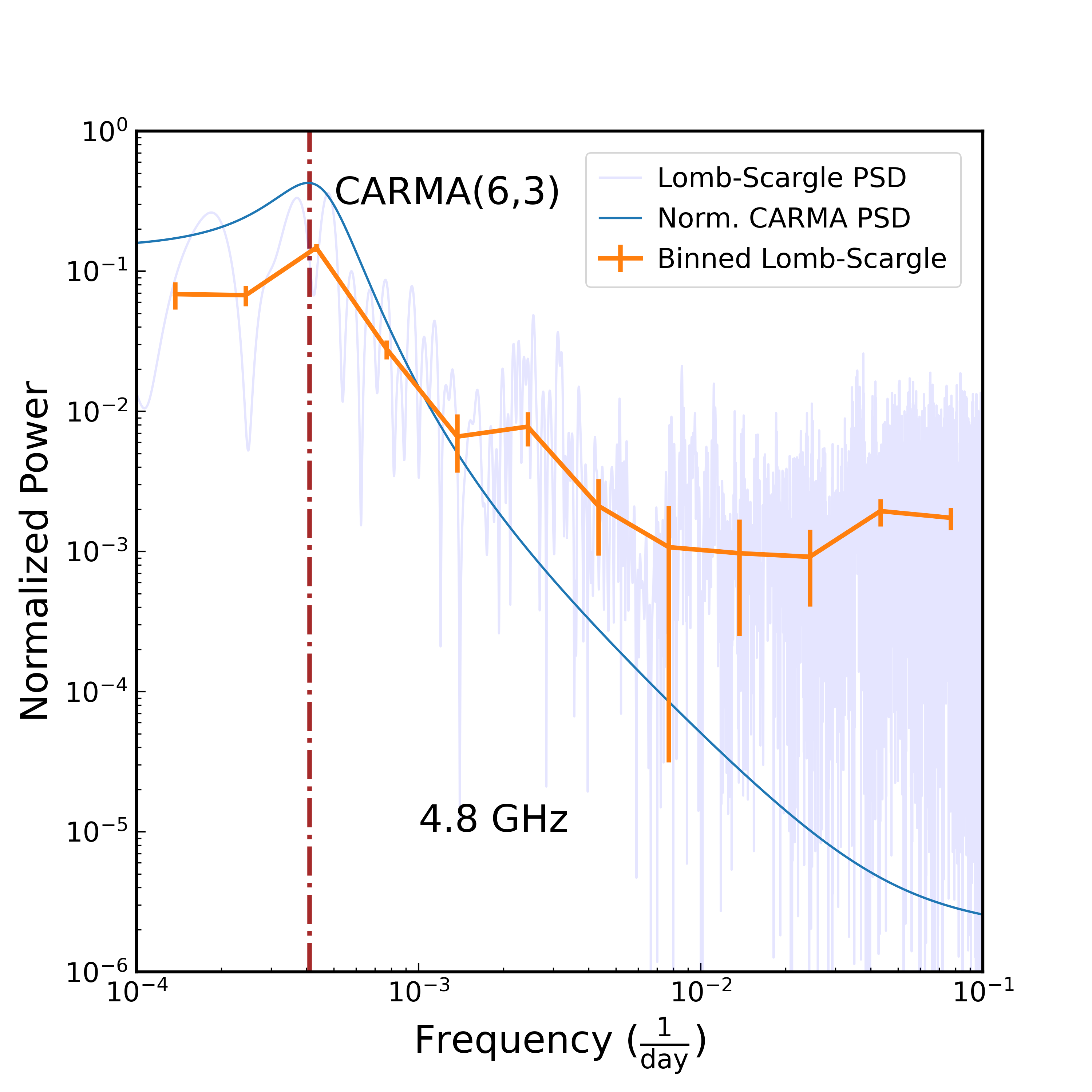}\hspace{-0.6cm} \includegraphics[scale=0.28]{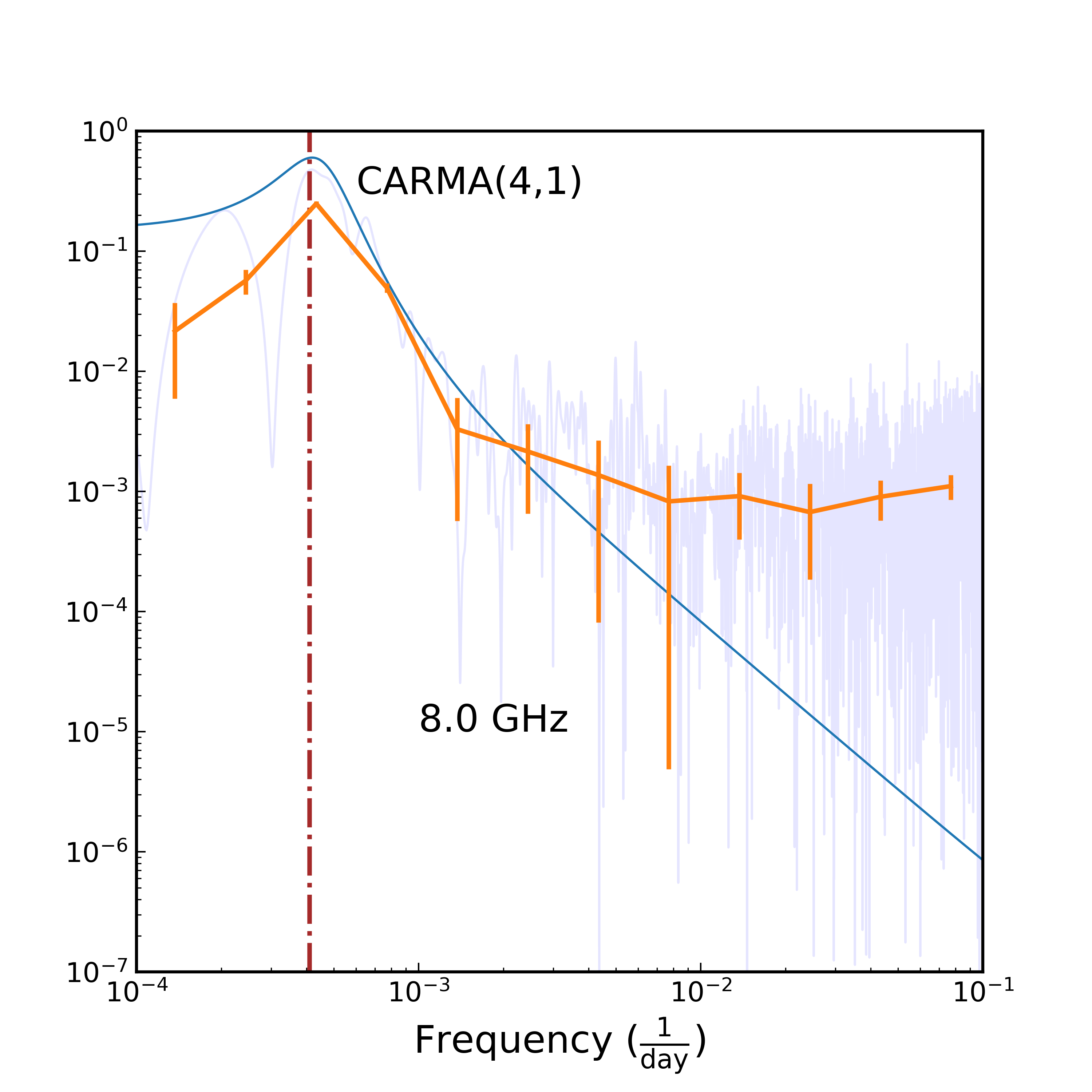}\hspace{-0.6cm}\includegraphics[scale=0.28]{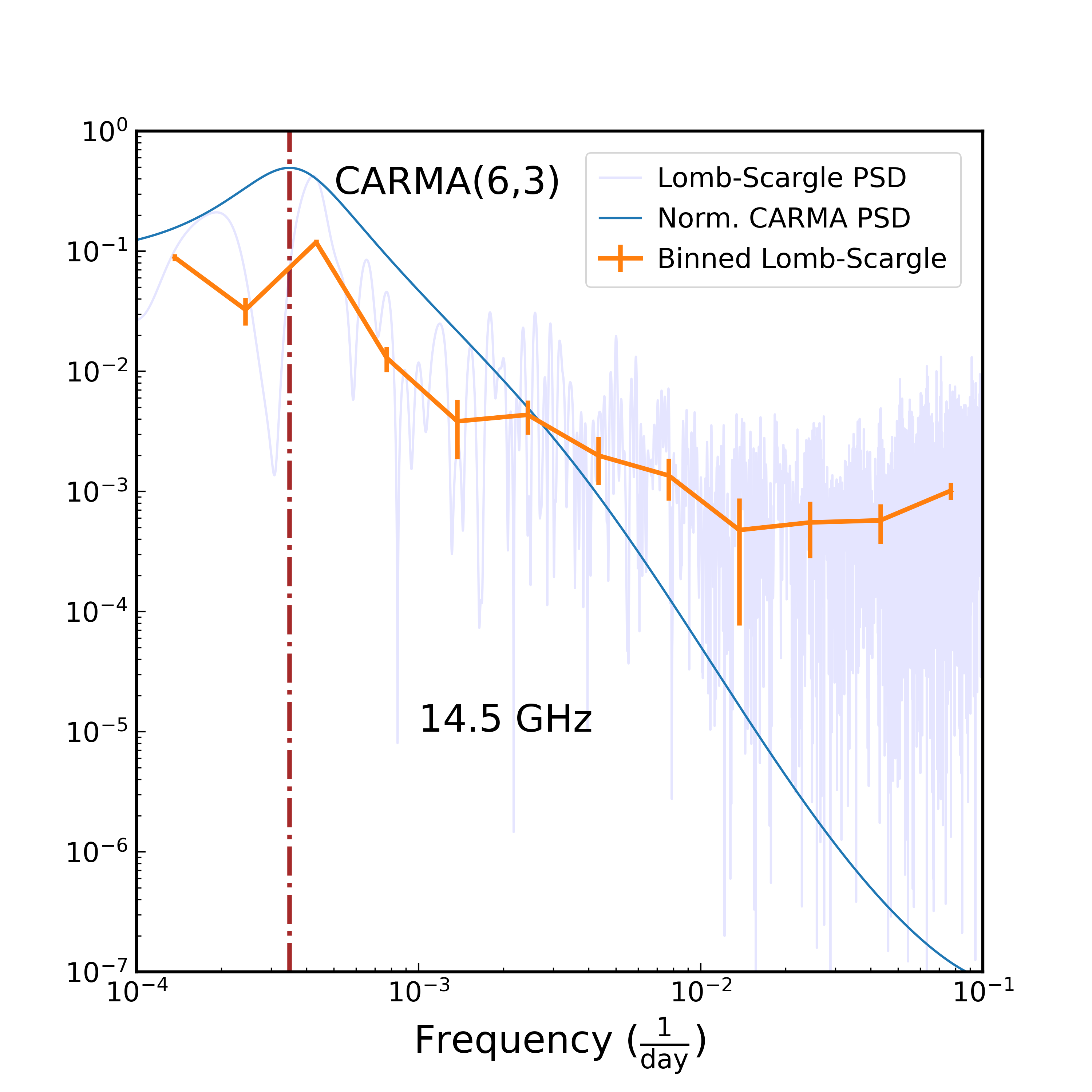}\\

\centering\includegraphics[scale=0.28]{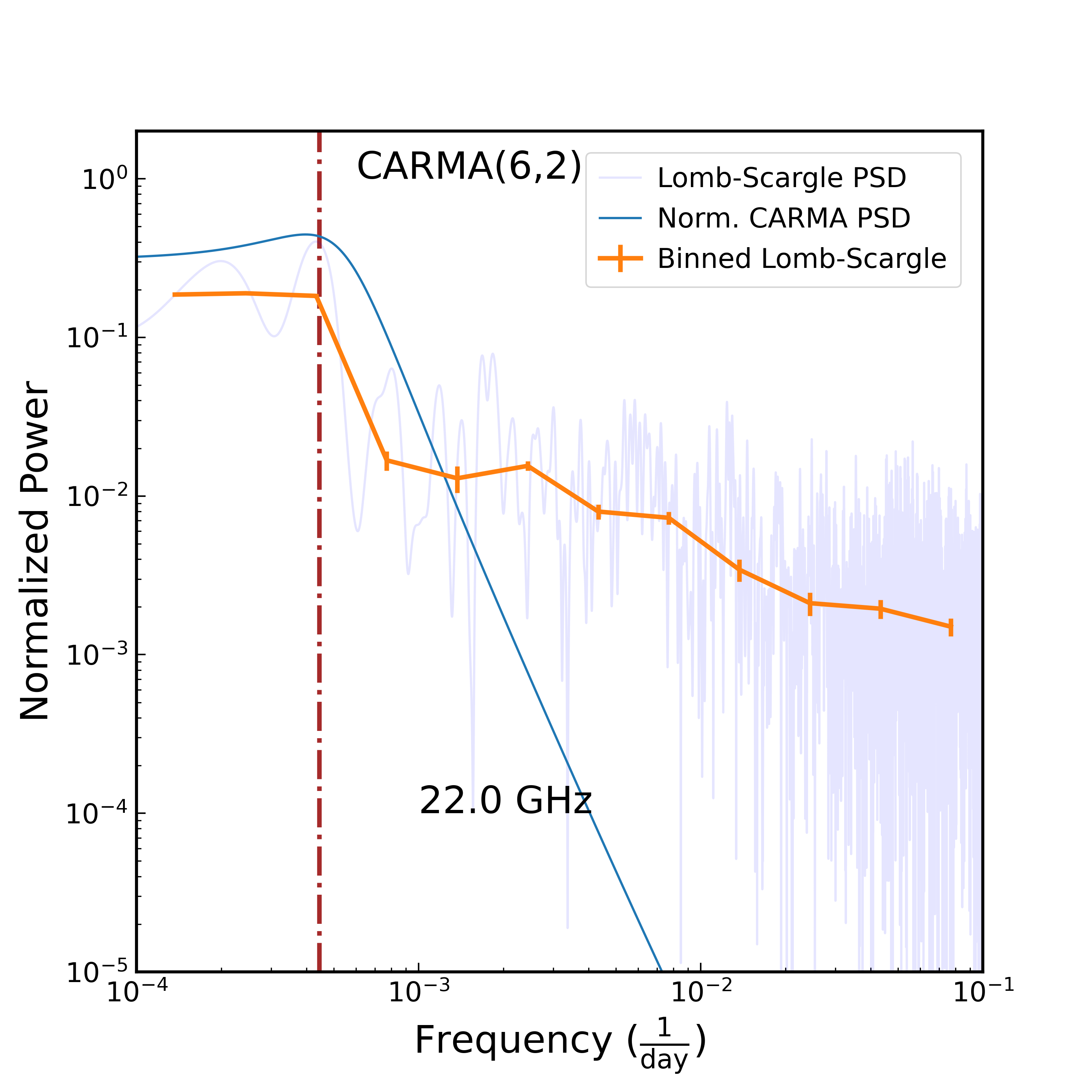}\hspace{-0.7cm} \includegraphics[scale=0.28]{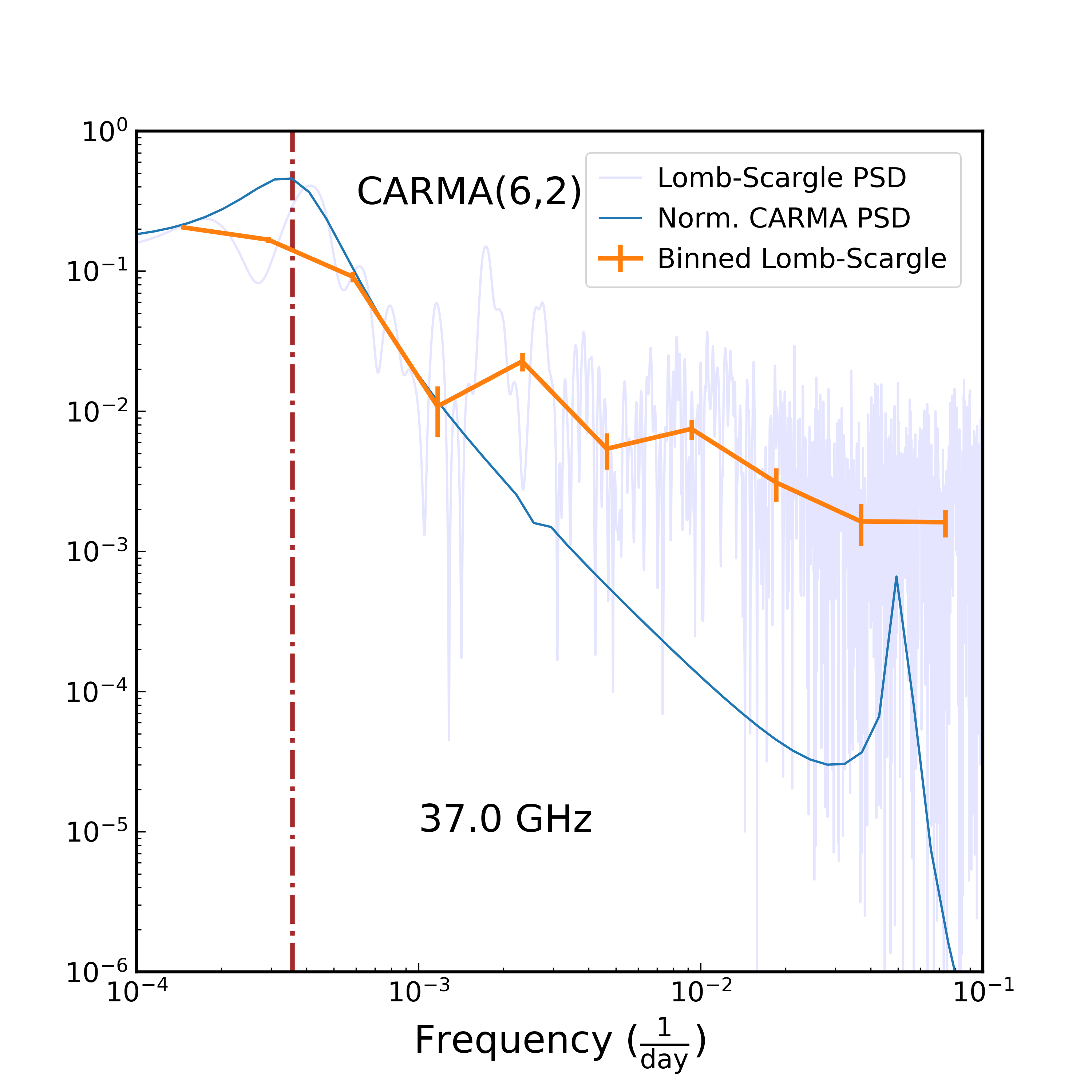}\\

\caption{The best-fit CARMA model PSD and binned Lomb-Scargle periodogram for the light curves  at 4.8, 8.0, 14.5, 22.0, and 37.0 GHz. The Lomb-Scargle periodogram is plotted for comparison. The brown dashed curve denotes the frequency at which the highest value of normalized power is detected. }
\end{figure*}\label{fig:carma}

\begin{figure*}[t]
\centering

\includegraphics[scale=0.35]{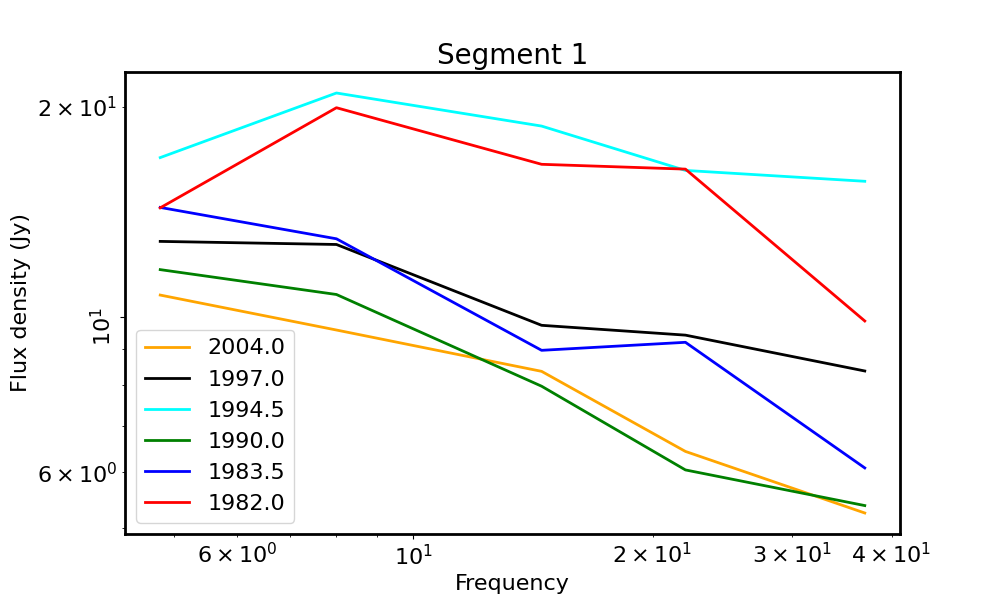}\includegraphics[scale = 0.35]{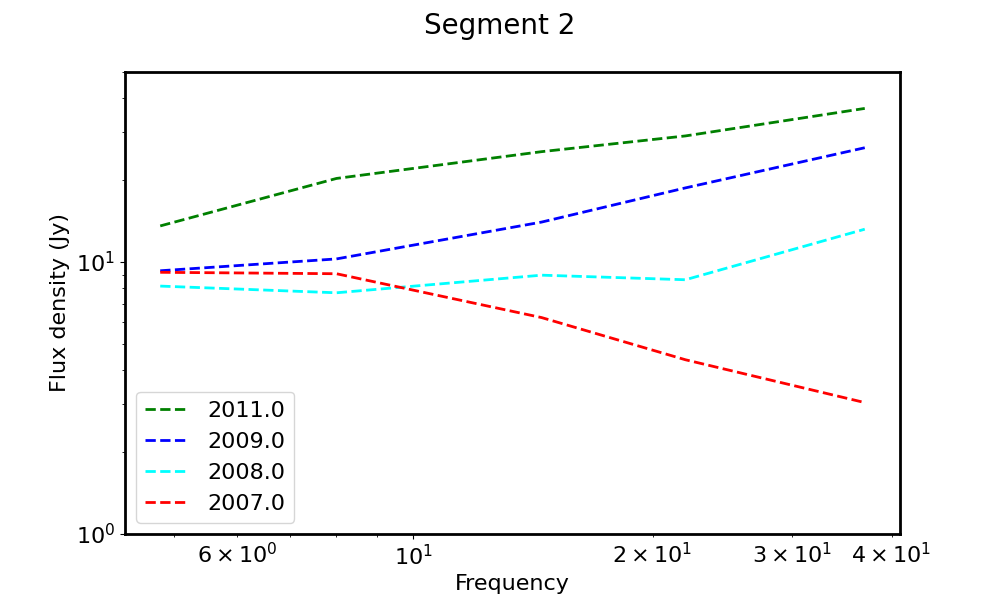}

\caption{Spectral energy distributions (SEDs) plotted at all five radio frequencies analyzed in this work. The left panel shows the SEDs during Segment 1 and the right panel shows the SEDs during Segment 2.}
\end{figure*}\label{fig:sed}

\noindent
In blazars, the non-thermal emission from jets and accretion disks dominates the cumulative emission in all EM bands. \black{Time-dependent changes in the fueling of the jet by the central engine (BH plus accretion disk) would affect the development and growth of the instabilities in the jets.}  And the origin of the emission at the radio frequencies employed in this work certainly comes from the jet which is strongly Doppler-boosted owing to its low inclination angle ($\sim 1.3^{\circ}$) \citep{2009A&A...494..527H}. 
Therefore, any quasi-periodicity observed is much more likely to be the result of internal jet processes than processes in the accretion disk. \\
\\
One likely origin of these long-period QPOs is the presence of a binary supermassive black hole system, as is almost certainly present in OJ 287 \citep[see][and references therein]{1996A&A...305L..17S,1996ApJ...460..207L, 2007A&A...462..547F,2017A&A...601A..52B, 2018MNRAS.478..359K, 2023ApJ...957L..11G, 2024ApJ...968L..17V}. In this now standard model for OJ 287, flares are produced when the secondary black hole smashes through the accretion disk around the primary. 
Alternatively in a binary model, the accretion rate onto one black hole can increase significantly when the other one comes in its proximity due to an elliptical orbit even if it does not impinge upon the disk. This increase in accretion rate could lead to the periodic increase in the flux \citep{2022ApJ...929..130W}. However, this model would be more important when the emission from the accretion disk exceeds that from the jet, which is unlikely in the case of blazars and particularly unlikely for radio emission. 
The orbital motion of the two black holes can naturally yield periodic fluctuations in observed radio flux from the change in Doppler factor caused by variation in the observation angle to the jet \citep[e.g.][]{2006A&A...453..817V, 2007ChJAA...7..364Q}. Recently, \citet{2022ApJ...926L..35O} applied this model to explain the quasi-periodicity of $\sim$ 1700 days in the blazar PKS 2131$-$021. So, the binary supermassive black hole hypothesis provides a plausible explanation for the $\sim$ 2000d QPO supported by our analysis. \\
\\
The wiggling of the jet could also be produced internally via Lens-Thirring precession occurring in the inner part of the accretion disk, from which the jet is presumably launched, due to relativistic frame-dragging \citep{1998ApJ...492L..59S, 2018MNRAS.474L..81L}. The Lens-Thirring precession model more naturally produces periodicity of the order of a few months. So, while this model is unlikely to explain the $\sim$2000d QPO it might produce the \black{putative} $\sim$600d one which is \black{apparently} detected in the light curves at higher radio frequencies (22.0 and 37.0 GHz). While it also may  be present at the lower frequencies, the broader peaks at lower frequencies could make it harder to detect.\\ 
\\
Another possible explanation of these QPOs is related to the internal helical structure in the jet \citep[see][and references therein]{2004ApJ...615L...5R, 2015ApJ...805...91M}. In this model, shocks generated in current-driven plasma are believed to propagate outwards toward the jets and interact with the toroidal magnetic field of the jet, leading to its distortion. These sudden changes affect the magnetic fields \textcolor{black}{around the black hole} and result in the periodic fluctuations manifested in the observed flux from the jet \citep[e.g.][]{1992A&A...255...59C, 2004ApJ...601..414L}. These periodic \black{changes} occur on the order of a few months to years and so could explain either the observed 2000d QPO or the \black{possible} 600d periodicity seen in this work. Magneto-hydrodynamic simulations suggest that some QPOs could be the result of quasi-periodic kinks arising in the jet due to the instability produced by distorted magnetic fields \citep[e.g.][]{2012MNRAS.423.3083M, 2020MNRAS.494.1817D}. However, if these fluctuations are due to kink instabilities, they would be by far the longest discovered so far, as kink-driven oscillations seem to persist for only days or weeks \citep[e.g.][]{2022Natur.609..265J,2024MNRAS.527.9132T,2024MNRAS.528.6608T}. \\
\\
\blue{Fig. \ 11} shows a series of spectral energy distributions (SEDs)  where the flux densities at all five radio frequencies are plotted at different times. The left plot shows the SEDs during Segment 1 and the right plot, during Segment 2. In Segment 1, the flux density usually declines monotonically with frequency, as is typical for radio synchrotron emission from a steady jet. But during the high emission states in 1982.0 and 1994.5,  the flux density peaks around 8 GHz. 
During these epochs the highest frequency emissions are past their peaks and already fading,  those around 8 GHz are near their peaks, while a 4.8 GHz the flux is still rising.  This can be understood if emission at lower frequencies is delayed and is broadened as it emerges from a larger region. 
This behavior of flux variation at different radio frequencies is expected with the standard shock-in-jet model \citep[e.g.,][]{1985ApJ...298..114M,1989ApJ...341...54H}. \\
\\
However, in Segment 2, the behavior of the SEDs is generally very  different. At the beginning of this period at 2007.0, when the source is dim, there is a continuation of the usual essentially monotonic decline of flux density with frequency.  Afterward, when the flaring process dominates, the SEDs shift toward a trend of increasing flux with rising frequency. This  behavior of the SEDs 
indicates that the physical process governing the strong flares in Segment 2 (2007 onwards) may be different from that of Segment 1. A likely scenario would be the presence of a knot emerging from the core and passing through the standing shock in the jet, as depicted in the Very Long Baseline Array (VLBI) image of 3C 454.3 at 43 GHz \citep{2008Natur.452..966M}. 
Such behavior, known as core-shift variability, has also been found in the VLBI images at different radio frequencies made between 2005 and 2010 \citep{2023A&A...672A.130C}, during the flares seen in Segment 2. 
\citet{2010ApJ...715..362J} analyzed multi-frequency light curves of this object during 2005---2008 as well as VLBA maps they made at frequent intervals and concluded that the variability predominantly arises from outward moving knots interacting with a stationary knot in the jet $\sim 0.6$ mas from the core. 
Recently, \citet{2024A&A...682A.154T} analyzed VLBI images at 43 and 86 GHz of this source during 2013--2017 and found that some superluminal features  abruptly disappeared at that stationary knot. They claimed that these peculiar kinematics can be explained by the presence of a bend in the jet at that characteristic location.\\ 
\\
One \black{interesting result of this work is the discovery of an apparent} $\sim$~600d signal, which is most prominent at higher frequencies. This signal becomes strong towards the middle of the observations and persists until the end of the observations of Segment 1. However, the flaring seen in Segment 2 apparently suppresses this signal.

\section*{ACKNOWLEDGMENTS}
\noindent
This work is supported by NASA grant number 80NSSC22K0741. This research is based on data from the University of Michigan Radio Astronomy Observatory, which was supported by the National Science Foundation, NASA, and the University of Michigan. UMRAO research was supported in part by a series of grants from the NSF (including AST-0607523 and AST-2407604) and by a series of grants from NASA, including Fermi G.I.\ awards NNX09AU16G, NNX10AP16G, NNX11AO13G, and NNX13AP18G. This publication makes use of data obtained at Mets{\"a}hovi Radio Observatory, operated by Aalto University in Finland. The various diligent observers of Aalto University are thankfully acknowledged. 


\end{document}